\documentclass[epj,nopacs]{svjour}
\pdfoutput=1
\usepackage[utf8]{inputenc}
\usepackage{color,graphicx,hyperref,cite,amsmath,slashed,arydshln,amssymb}
\hypersetup{colorlinks=true,linkcolor=blue,citecolor=magenta,filecolor=magenta,urlcolor=cyan}
\graphicspath{{figures/}}

\newcommand{\be}{\begin{equation}}
\newcommand{\ee}{\end{equation}}
\def\bsp#1\esp{\begin{split}#1\end{split}}

\newcommand{\madanalysis}{{\sc MadAnalysis~5}}

\newcommand{\delphes}{{\sc Delphes~3}}

\newcommand{\madgraph}{{\sc MG5\_aMC}}
\newcommand{\pythia}{{\sc Pythia~8}}
\newcommand{\rivet}{{\sc Rivet}}
\newcommand{\rooot}{{\sc Root}}
\newcommand{\fastjet}{{\sc FastJet}}

\def\ie{{\it i.e.}}
\def\etc{{\it etc.}}

\begin{document}
\title{Simplified fast detector simulation in \madanalysis}
\author{
  Jack Y. Araz\inst{1,2}\thanks{\color{blue}jack.araz@glasgow.ac.uk},
  Benjamin Fuks\inst{3,4}\thanks{\color{blue}fuks@lpthe.jussieu.fr} and
  Georgios Polykratis\inst{3}\thanks{\color{blue}giopolykra@gmail.com}
}

\institute{
    School of Physics \& Astronomy, University of Glasgow, United Kingdom
  \and
    Concordia University 7141 Sherbrooke St. West, Montr\'{e}al, QC,
    Canada H4B 1R6
  \and
    Laboratoire de Physique Th\'eorique et Hautes Energies (LPTHE),
    UMR 7589, Sorbonne Universit\'e et CNRS, 4 place Jussieu,
    75252 Paris Cedex 05, France
  \and
    Institut Universitaire de France, 103 boulevard Saint-Michel, 75005 Paris,
    France
}

\date{}

\abstract{
We introduce a new simplified fast detector simulator in the
\madanalysis\ platform. The {\sc Python}-like interpreter of the programme has
been augmented by new commands allowing for a detector parametrisation through
smearing and efficiency functions. On run time, an associated C++ code is
automatically generated and executed to produce reconstructed-level events. In
addition, we have extended the \madanalysis\ recasting infrastructure to support
our detector emulator, and we provide predefined LHC detector configurations. We
have compared predictions obtained with our approach to those
resulting from the usage of the \delphes\
software, both for Standard Model processes and a few new physics signals.
Results generally agree to a level of about 10\% or better, the
largest differences in the predictions stemming from the different
strategies that are followed to model specific detector effects. Equipped with
these new functionalities, \madanalysis\
now offers a new user-friendly way to include detector effects when analysing
collider events, the simulation of the detector and the analysis being both
handled either through a set of intuitive {\sc Python} commands or directly
within the C++ core of the platform.
}

\titlerunning{Simplified fast detector simulation in \madanalysis}
\authorrunning{J.Y.~Araz {\it et al.}}

\maketitle

\section{Introduction}\label{sec:intro}
The discovery of the last missing particle of the Standard Model at the Large
Hadron Collider (LHC) at CERN has opened a new era in our understanding of the
fundamental laws of nature. However, the concrete mechanism behind electroweak
symmetry breaking is still today a mystery and there is no sign of phenomenon
beyond the Standard Model, despite the wealth of data currently available. As a
consequence, the experimental LHC search results are interpreted as stronger and
stronger constraints on a large set of new physics models, those constraints
being obtained by comparing associated predictions with data.

Those reinterpretations can be achieved in two ways. First, they can rely on
experimental data from which the detector effects have been unfolded, \ie\ by
considering LHC data as observed by a perfect detector with an infinite
resolution and an ideal calibration. This requires an excellent understanding of
the background and is a complex, ill-defined and time-consuming problem, as
there is no unique solution to the inversion of the convolution of the detector
response~\cite{Spano:2013nca}. Consequently, reinterpretations are usually
performed by adopting a second approach in which the detector effects are
included, or folded forward, in the simulation of the new physics signals to be
confronted to data. This folding is expected to appropriately capture the impact
of the inner working of the detector and the inefficiency of the reconstruction
of the event record. This has the advantage of being computationally much more
acceptable.

The most accurate forward folding method relies on the implementation of the
exact details of the detector functioning in a framework based on the
{\sc Geant~4} package~\cite{Agostinelli:2002hh}. The latter allows for the
modelling of the detector geometry and material interactions, from which one
could then deal with the simulation of the electronic response of the detector.
As a final step, it is required to reproduce the impact of the
reconstruction details associated with any given experiment, which yields
appropriate definitions for the various physics objects used in an analysis.
This is, however, in practice not achievable by anyone outside the
collaborations, by virtue of the lack of publicly available information. In
addition, even if feasible, running such a chain of tools for the plethora of
new physics model potentially interesting is likely to be computationally
unfeasible, as this requires several minutes of computing time for a single
simulated event.

The resolution, reconstruction and identification efficiencies corresponding to
all final-state objects relevant for a physics analysis are however often
publicly available under the form of functions of standard object properties
(like the transverse momentum or pseudo-rapidity). This, therefore, opens the
door to a well-motivated and computationally much more efficient approach, as is
implemented in packages like the \delphes\ software~\cite{deFavereau:2013fsa} or
the \rivet\ toolkit~\cite{Buckley:2010ar}.

In \delphes, the simulation of the detector approximates the steps followed
in a {\sc Geant}-based approach. It relies on approximate experiment
geometries and particle propagation models, that are combined with tabulated
reconstruction and identification efficiencies to yield the reconstructed
physics objects to use in an analysis.
The latter are obtained from hadron-level events clustered through
one of the different algorithms available from the \fastjet\
package~\cite{Cacciari:2011ma}, together with potentially
more complex techniques, such as particle-flow
and energy-flow methods.
Moreover, the programme allows for the
optional simulation of pile-up. In \rivet, a lighter approach is
implemented~\cite{Buckley:2019stt}. Effective transfer functions, including the
smearing of the object kinematics properties and reconstruction efficiencies,
are used to connect the Monte Carlo representation of any physics object (\ie\
at the truth level) to their reconstructed representation.

In this paper, we report on the design of an extension of the capabilities of
the \madanalysis\ framework~\cite{Conte:2012fm,Conte:2014zja,Conte:2018vmg} to
handle the simulation of the response of a detector. \madanalysis\ is a general
platform for beyond the Standard Model phenomenology. It can deal both with the
development of an analysis of any given collider signal (together with its
associated background) via a user-friendly {\sc Python}-based command-line
interface and a developer-friendly C++ core. \madanalysis\ can also be
used for the (automated) reinterpretation of existing LHC results~\cite{%
Dumont:2014tja,Araz:2019otb}.

The response of a typical detector can be emulated in \madanalysis\ in two ways.
The user has first the option to rely on the interface of the code to the
\fastjet\ package~\cite{Cacciari:2011ma}. Whilst fast and efficient,
this way of proceeding is restricted to the simulation of a detector with an
infinite resolution. It indeed leads to the sole application of a jet clustering
algorithm (as available from \fastjet) to reconstruct events, possibly
together with flat parametrisations to model some specific reconstruction
effects like $b$-tagging or tau-tagging. Nevertheless, this procedure is useful
for studies dedicated to a particular effect or at the Monte Carlo truth level.

As a simple application of a jet algorithm is known to be quite unrealistic in
many contexts for which detector effects matter, \madanalysis\ has been
interfaced to the \delphes\ package to enable a more detailed simulation
of the detector effects. This interface drives the run of \delphes, and
additionally switches on the usage of new modules that are specific to
\madanalysis. These are for instance related to a better handling of object
isolation or the skimming of the output \rooot\ file. The price to
pay to employ \delphes\ instead of simply \fastjet\ is obviously a much
slower run of the code, as stemming from the larger complexity of \delphes\ with
respect to \fastjet.

In the following, we detail several improvements that have been added to the
\madanalysis\ interface to \fastjet. These allow the user to
include a light\-weight, therefore computationally cheap, and realistic detector
simulation when Monte Carlo event reconstruction (with \fastjet) is in order.

Thanks to those developments, \madanalysis\ offers now the flexibility to rely
on any specific set of transfer functions to effectively map the Monte Carlo
objects to their reconstructed counterparts (\ie\ reconstruction efficiencies,
tagging efficiencies and the corresponding mistagging rates). Moreover, the
properties of the different objects can be smeared to mimic detector
resolution degradations.
\madanalysis\
becomes thus the only publicly available high-energy physics package that offers
the user the choice of either relying on lightweight smearing and efficiency
functions or the heavier \delphes\ framework to model the response of a typical
high-energy physics detector, easing hence the potential comparison of the pros
and cons and of two methods within a single platform.
Moreover, thanks to the \madanalysis\ {\sc Python} interpreter and the
associated intuitive meta-language, the parametrisation of the detector is
user-friendly and the generation of the corresponding C++ code (either when a
\delphes-based or transfer-function-bases simulation is used), its compilation
and running happen behind the scenes in a fully automated fashion.
Our changes in the C++ core of the programme have been
combined with new functionalities at the level of the {\sc Python} command-line
interface, so that any user can straightforwardly implement his/her own detector
pa\-ra\-me\-tri\-sa\-ti\-on in a simple and user-friendly manner. In addition,
the code is shipped with predefined sets of commands allowing to automatically
load detector configurations associated with the ATLAS and CMS detectors.

The new features introduced in this work are available from version 1.8.51 of
the code, that can
can be downloaded from \href{https://launchpad.net/madanalysis5}{\sc LaunchPad}%
\footnote{See the webpage \url{https://launchpad.net/madanalysis5}.}. As said
above, this release of the programme allows for the simulation of the
impact of a detector in a way that combines realism and efficiency. In addition,
it allows for dedicated studies singling out specific detector effects, that
could potentially be applied directly at the particle level. The
information included in this paper is summarised on the 
\href{http://madanalysis.irmp.ucl.ac.be/wiki/SFS}{\madanalysis\ website}%
\footnote{See the webpage \url{http://madanalysis.irmp.ucl.ac.be/wiki/SFS}.}.

The rest of this paper has been prepared as follows. In section~\ref{sec:reco},
we briefly review how to reconstruct a sample of Monte Carlo events in
\madanalysis\ with the help of its interface to \fastjet. In section~\ref{%
sec:sfs}, we detail the new features that have been developed in order to allow
the user to combine event reconstruction with a fast and realistic detector
simulation. In section~\ref{sec:validation}, we compare predictions obtained
with our new method with those arising from the usage of \delphes. We moreover
quantitatively assess the differences between the results returned by both
detector simulators relatively to the Monte Carlo truth. We then present, in
section~\ref{sec:pad}, how the new \madanalysis\ fast and simplified detector
simulator can be used for the reinterpretation of the results of two specific
LHC analyses. In this context, a comparison with a more usual method relying on
\delphes\ is performed. We summarise our work and conclude in
section~\ref{sec:conclusion}.

\section{Event reconstruction with \madanalysis}\label{sec:reco}

\subsection{Running \fastjet\ from \madanalysis}
\label{sec:fjrun}
\begin{table*}
  \centering
  \renewcommand{\arraystretch}{1.4}
  \setlength\tabcolsep{8pt}
  \begin{tabular}{c | c c c | c c c c}
    Algorithm & {\tt ptmin} & {\tt exclusive\_id} & {\tt radius} &
     {\tt p} & {\tt overlap} & {\tt npassmax} & {\tt input\_ptmin}\\ \hline
      \texttt{kt}          & 5 & {\tt true} & 1 & - & -   & - & - \\
      \texttt{cambridge}   & 5 & {\tt true} & 1 & - & -   & - & - \\
      \texttt{antikt}      & 5 & {\tt true} & 1 & - & -   & - & - \\
      \texttt{genkt}       & 5 & {\tt true} & 1 & 1 & -   & - & - \\
      \texttt{siscone}     & 5 & {\tt true} & 1 & - & 0.5 & 0 & 0 \\
  \end{tabular}
  \caption{\it Jet algorithms available in \madanalysis, shown together with the
    properties than can be changed by the user on run time. The default values
    are indicated. The jet radius default value of 1 has been replaced by 0.4
    in \madanalysis\ releases posterior to v1.9.11.}
  \label{tab:fj_props}
\end{table*}

Thanks to its interface to \fastjet, \madanalysis\ allows for the reconstruction
of hadron-level Monte Carlo events through the application of a jet-clustering
algorithm. To proceed, the programme has to be started in the
reconstructed mode,
\begin{verbatim} ./bin/ma5 -R\end{verbatim}
and the \fastjet\ package has to be present on the user system. If this is not
the case, it is sufficient to type, in the \madanalysis\ command-line interface,
\begin{verbatim} install fastjet\end{verbatim}
to trigger a local installation of \fastjet, in the subfolder
{\tt tools/fastjet}.

The \madanalysis\ run has then to be configured following the specifications of
the user. First, one must switch event reconstruction on by typing
\begin{verbatim} set main.fastsim.package = fastjet\end{verbatim}
This turns on the usage of \fastjet\ for event reconstruction and indicates to
the code that the input hadron-level event sample(s) are encoded following
the {\sc HepMC} event file format~\cite{Dobbs:2001ck}\footnote{The deprecated
\href{https://cdcvs.fnal.gov/redmine/projects/heppdt/wiki/Notes\_about\_StdHep}
{{\sc StdHep} format} is still supported in \madanalysis\ and can
possibly be used as well.}.

By default, jets are reconstructed by making use of the anti-$k_T$
algorithm~\cite{Cacciari:2008gp}, with a radius parameter set to
$R=1$ for \madanalysis\ releases prior to version 1.9.11 and $R=0.4$
for more recent versions\footnote{The $R=1$ default choice is a
remnant of the early days of the programme.}.
This behaviour can be modified by typing in
\begin{verbatim} set main.fastsim.algorithm = <algo>
 set main.fastsim.<property>= <value>\end{verbatim}
where \verb|<algo>| represents the keyword associated with the adopted jet
algorithm, and {\tt <property>} generically denotes any of its property.
\madanalysis\ can employ the longitudinally invariant
$k_T$ algorithm~\cite{Catani:1993hr,Ellis:1993tq} ({\tt kt}), the
Cambridge/Aachen algorithm~\cite{Dokshitzer:1997in,Wobisch:1998wt} ({\tt
cambridge}), the anti-$k_T$ algorithm ({\tt antikt}), the generalised $k_T$
algorithm~\cite{Cacciari:2011ma} ({\tt genkt}), as well as
the seedless infrared-safe cone algorithm~\cite{Salam:2007xv}
({\tt siscone})\footnote{While other cone algorithms are interfaced, they are
not infrared-safe and will therefore not be further discussed. We refer to
ref.~\cite{Conte:2018vmg} for more information.}. We refer to
table~\ref{tab:fj_props} for the list of available jet algorithms, the
corresponding options and their default values.

Among all the jet algorithm properties that can be tuned, three of them are
common to all algorithms. The user can define a transverse momentum threshold
({\tt ptmin}) so that any softer jet is filtered out, and fix the jet radius
parameter $R$ ({\tt radius}) that dictates how distant the constituents of a
given jet can be. Moreover, he/she can decide whether the algorithm should be
exclusive or inclusive ({\tt exclusive\_id}), \ie\ whether the leptons and
photons originating from hadron decays have to be included in their respective
collections in addition to be considered as constituents of the reconstructed
jets ({\tt exclusive\_id = false}), or not ({\tt exclusive\_id = true}).

Furthermore, the generalised $k_T$ algorithm involves a distance measure
depending on a continuous parameter $p$ ({\tt p}), and the siscone algorithm
depends on the fraction of overlapping momentum above which two protojets are
combined ({\tt overlap}), on the maximum number of passes the algorithm should
be carried out ({\tt npassmax}), and on a threshold driving the removal of
too soft objects ({\tt input\_ptmin}). Those protojets, on which jet
reconstruction relies in general, are either final-state hadrons or objects that
have been already combined. For all algorithms, the combination process obeys to
the $E$-scheme~\cite{Cacciari:2011ma}, \ie\ the momentum of the combined object
equals the sum of the initial momenta.

In practice, the code starts by filtering the input particles (as provided in
the event record), restraining the analyses list to all visible
final-state particles. This corresponds to the entire set of objects not
explicitly tagged as invisible. The tagging of any object as invisible can be
achieved by typing in
\begin{verbatim} define invisible = invisible <pdg-code> \end{verbatim}
where the \verb+<pdg-code>+ value corresponds to the Particle Data Group (PDG)
identifier~\cite{Tanabashi:2018oca} of a new invisible state. The above command
results in adding this new code to the list of invisible particles stored in the
\verb+invisible+ container, which includes by default the Standard Model
neutrinos and antineutrinos, as well as the supersymmetric lightest neutralino
and gravitino.

Similarly, the user can inform the code about the existence of a new
strongly-interacting particle, which hence participates to the hadronic activity
in the event and has to be accounted for by the clustering algorithm. The
information is provided by superseding the definition of the \verb+hadronic+
container,
\begin{verbatim} define hadronic = hadronic <new-pdg-code>\end{verbatim}

After having defined the characteristics of the reconstruction, event files
have to be imported, either one by one or simultaneously by using wildcards.
This is achieved through the standard command
\begin{verbatim} import <path-to-hepmc-files> as <set>\end{verbatim}
where the user-defined label \verb|<set>| allows one to group several event
files
(assumed to describe the same physics process) into a single set. This line can
be repeated as much as needed. The reconstructed events are  saved on disk and
stored in an event file encoded in the LHE event format~\cite{Boos:2001cv,
Alwall:2006yp} by typing, in the interpreter,
\begin{verbatim} set main.outputfile = <name-of-an-LHE-file> \end{verbatim}
where \verb+<name-of-an-LHE-file>+ is a filename carrying an \verb+.lhe+ or
\verb+.lhe.gz+ extension. This file will be created in the
\verb+<wdir>/Output/SAF/<set>/lheEvents0_0+ directory during the \madanalysis\
run (\verb+<wdir>+ denoting the run working directory).

The clustering is finally started after the
\begin{verbatim} submit\end{verbatim}
command is entered. This results in the automated generation of a C++ code
representing the defined reconstruction process, its compilation and its
execution on the input event sample(s).

\subsection{Crude detector simulation}
\label{sec:crudesim}

Already in its earlier versions (\ie\ without the novelties introduced in this
paper), \madanalysis\ gives the option to include basic detector effects in the
reconstruction process, such as simple tagging and mistagging efficiencies.

\begin{table}
  \centering
  \renewcommand{\arraystretch}{1.4}
  \setlength\tabcolsep{8pt}
  \begin{tabular}{c | c c}
    Properties& Default value & type\\ \hline
      \verb+matching_dr+ & 0.5         & Float\\
      \verb+efficiency+  & 1           & Float in $[0, 1]$\\
      \verb+exclusive+   & \verb+true+ & Boolean\\
      \verb+misid_cjet+  & 0           & Float in $[0, 1]$\\
      \verb+misid_ljet+  & 0           & Float in $[0, 1]$\\
  \end{tabular}
  \caption{\it Properties defining the crude $b$-tagging algorithm available
    in older versions of \madanalysis, shown together with their default
    values. These properties can be modified by typing in the interpreter
    {\tt set main.fastsim.bjet\_id.<property> = <value>}.}
  \label{tab:crude_btag}
\end{table}

\begin{table}
  \centering
  \renewcommand{\arraystretch}{1.4}
  \setlength\tabcolsep{8pt}
  \begin{tabular}{c | c c}
    Properties& Default value & type\\ \hline
      \verb+efficiency+  & 1           & Float in $[0, 1]$\\
      \verb+misid_ljet+  & 0           & Float in $[0, 1]$\\
  \end{tabular}
  \caption{\it Properties defining the crude tau-tagging algorithm available
    in older versions of \madanalysis, shown together with their default
    values. These properties can be modified by typing in the interpreter
    {\tt set main.fastsim.tau\_id.<property> = <value>}.}
  \label{tab:crude_tautag}
\end{table}

The platform first allows for the implementation of a simple $b$-tagging
procedure, relying on flat efficiencies and mistagging rates. The decision
behind the (mis)tagging of any given jet involves the angular distance between
the reconstructed object and a true $B$-hadron (as read from the input event).
Each property defining this procedure is entered in the interpreter as follows,
\begin{verbatim} set main.fastsim.bjet_id.<property> = <value>\end{verbatim}
their list and default values being given in table~\ref{tab:crude_btag}.
Behind the scenes, the algorithm matches each $B$-hadron present in the analysed
event to all reconstructed jets lying at an angular distance smaller than a
threshold fixed by the user (\verb+matching_dr+). Those jets are then considered
as $b$-tagged with a probability entered by the user (\verb+efficiency+).
$b$-jet identification can be restricted to the closest jet
(\verb+exclusive = true+), or to all matched jets (\verb+exclusive = false+).
The mistagging of charm and light jets as $b$-jets is performed similarly,
the decision being taken following flat probabilities that are provided by the
user (\verb+misid_cjet+ and \verb+misid_ljet+ respectively).

Next, a simple tau identification procedure can be employed, its properties
being set by typing, in the interpreter,
\begin{verbatim} set main.fastsim.tau_id.<property> = <value>\end{verbatim}
The list of available options, together with their default values, is given in
table~\ref{tab:crude_tautag}. The user has the possibility to fix the
probability with which a jet originating from the hadronic decay of a tau lepton
will be correctly tagged as a hadronic tau object (\verb+efficiency+), as well
as the probability with which a light jet will be incorrectly tagged as a
hadronic tau (\verb+misid_ljet+).

\section{A simplified and realistic fast detector simulator}\label{sec:sfs}

\subsection{Generalities}\label{sec:design_aim}

In the present work, we have improved the way in which detector effects can be
included when the interface of \madanalysis\ with \fastjet\ is used for event
reconstruction. Our modifications allow for the post-processing of the \fastjet\
output to model detector effects on the basis of transfer functions. Those
functions are provided by the user directly in the {\sc Python} interpreter and
could depend on various object kinematical properties. When the C++ code is
generated, the transfer functions are converted into a new C++ module that is
called at the end of the \fastjet\ run. This enables the modification of the
properties of the reconstructed objects according to various experimental
biases. In contrast to relying on a complex programme like \delphes\ that acts
at the hadronic level, our setup acts at the time of the reconstruction, which
results in a faster event reconstruction process and lighter output files.

\begin{table}
  \centering
  \renewcommand{\arraystretch}{1.4}%
  \setlength\tabcolsep{8pt}
 \begin{tabular}{|p{8.2cm}|}\hline
  \verb?define reco_efficiency <obj> <func> [<dom>]?\\ \vspace{-3.5mm}
    Defines the efficiency associated with the reconstruction of an object
    \verb+<obj>+. The efficiency is provided as a piecewise function whose each
    component \verb+<func>+ has an optional  domain of definition \verb+<dom>+.
    \\
  \verb?define smearer <obj> with <obs> <func> [<dom>]?\\ \vspace{-3.5mm}
    Defines the standard deviation of a Gaussian of vanishing mean that is
    relevant for the smearing of the property \verb+<obs>+ of the object
    \verb+<obj>+. The width of the Gaussian is provided as a piecewise function
    whose each component \verb+<func>+ has an optional domain of definition
    \verb+<dom>+. \\
  \verb?set main.fastsim.jetrecomode = <value>?\\ \vspace{-3.5mm}
  Allows to switch between jet-based jet smearing (\verb+<value> = jets+,
     default) and constituent-based jet smearing
     (\verb+<value> = constituents+).\\
  \verb?define tagger <obj> as <reco> <func> [<dom>]?\\ \vspace{-3.5mm}
    Defines the efficiency of tagging the Monte Carlo truth object \verb+<obj>+
    as a reconstructed \verb+<reco>+ object. The efficiency is provided as a
    piecewise function whose each component \verb+<func>+ has an optional domain
    of definition \verb+<dom>+.\\
  \verb?display reco_efficiency?\\ \vspace{-3.5mm}
  \verb?display smearer?\\ \vspace{-5.0mm}
  \verb?display tagger?\\ \vspace{-6.5mm}
    Displays the different modules of the implemented fast detector simulation.
    \\
  \hline
  \end{tabular}
  \caption{\it The subroutines allowing for the definition of a simplified fast
    detector simulation in \madanalysis. Reconstruction efficiencies, smearing
    and object identification are further detailed in sections~\ref{%
    sec:reco_eff}, \ref{sec:smear} and \ref{sec:tag} respectively.
    \label{tab:module}}
\end{table}

We consider three classes of effects. Firstly, the kinematical properties of any
given reconstructed objects could be smeared to account for the detector
resolution. Secondly, each object has a given probability of being effectively
reconstructed, depending on its kinematics. Finally, object identification can
be more or less successful, depending again on the kinematics, and leads to a
potential misidentification. Whilst the latter effects could already be
accounted for in the previous version of the code (see section~\ref{%
sec:crudesim}), it was not possible to include any dependence on the object
kinematics. This limitation has been alleviated.

In order to handle functions at the level of the com\-mand-line interface, we
rely on abstract syntax trees to decode the information provided by the user and
store it internally. Those trees can then be converted into C++ (or any
other programming language) and merged with the code generated by the
\madanalysis\ interpreter, that is then compiled and run on the input events.
This allows for a very flexible definition of any transfer function. The latter
is allowed to depend on any of the observables supported by \madanalysis,
the complete list of them being available from the manual~\cite{Conte:2012fm} or
the normal mode reference card (see App.~A of Ref.~\cite{Conte:2018vmg}).
Moreover, they can involve a variety of standard mathematical functions like
trigonometric, cyclometric or hyperbolic functions.

In the rest of this section, we discuss reconstruction efficiencies, smearing
and object identification in sections~\ref{sec:reco_eff}, \ref{sec:smear} and
\ref{sec:tag} respectively. The main commands to be typed in the command-line
interface are collected in table~\ref{tab:module}. At any time, the user can
display the currently implemented detector simulator modules by means of the
\verb+display+ command (see table~\ref{tab:module}).

As the event file import,
the generation of the working directory and the execution of the code are
unchanged, we refer to section~\ref{sec:fjrun}. Details on the
usage of the \madanalysis\ simplified fast detector simulation in the
expert mode of the programme are provided in section~\ref{sec:expert}, and
section~\ref{sec:lhc} describes how to make use of the CMS and ATLAS detector
parametrisation built in the \madanalysis\ platform.

\subsection{Reconstruction efficiencies}\label{sec:reco_eff}

The granularity of a typical high-energy physics detector, together with the
lack of precision in its data acquisition system, implies that it is not always
possible to fully reconstruct every single object that leaves hits in it.
This can be embedded in the \madanalysis\ machinery by defining
reconstruction efficiencies from the interpreter. The code will then
generate on run time a probability distribution indicating whether an object
should be reconstructed, according to its properties.

This is achieved by means of the \verb+define+ keyword,
\begin{verbatim} define reco_efficiency <obj> <func> [<dom>]\end{verbatim}
the first argument (\verb+reco_efficiency+) indicating that one deals with the
definition of a new reconstruction efficiency. In the above syntax, \verb|<obj>|
stands for the object under consideration, \verb|<func>| for the functional
form of the efficiency and \verb+<dom>+ consists in an optional attribute
relevant for piecewise functions. In the latter case, it is indeed necessary to
define, through (in)equalities, the domain of application of each piece of the
full function. In the case where the user would input an ill-defined function
for which the different kinematics subdomains are not disjoint (which is not
recommended), the code considers a weighted sum of the corresponding efficiency
functions. In addition, if this domain \verb+<dom>+ is not provided, the
efficiency is understood as applicable over the entire kinematical regime.

The list of available reconstructed objects is given in table~\ref{tab:sfs_objs}
and can be referred to either through a dedicated label, or through a so-called
`generalised PDG code'. The latter extends the traditional usage of PDG codes in
high-energy physics software in the sense that the code refers here to
reconstructed objects instead of the corresponding Standard Model particles. As
shown in the table, reconstruction efficiencies can be defined for jets
(\verb|j|), hadronic taus (\verb|ta|), electrons (\verb|e|), muons
(\verb|mu|) and photons (\verb|a|). At this stage of the detector simulation,
the distinction between heavy-flavour and light jets has not been implemented.
This is left for the particle identification module (see section~\ref{sec:tag}).
Moreover, those object definitions always refer to objects that are
not originating from hadronic decay processes. Any particle that would be
related to a hadronic decay is instead used as input for jet clustering.

\begin{table}
  \centering
  \renewcommand{\arraystretch}{1.4}
  \setlength\tabcolsep{8pt}
  \begin{tabular}{c | c c}
    Object & PDG code & Label \\  \hline
     Electron  & 11  & \verb+e+  \\
     Muon      & 13  & \verb+mu+ \\
     Tau       & 15  & \verb+ta+ \\
     Jet       & 21  & \verb+j+  \\
     Photon    & 22  & \verb+a+  \\
    \hdashline
     $c$-jet   & 4   & \verb+c+  \\
     $b$-jet   & 5   & \verb+b+  \\
  \end{tabular}
  \caption{\it List of reconstructed objects supported by the code. They are
    given
    together with their generalised PDG code and the corresponding \madanalysis\
    label through which they can be referred to in the code. $b$-jets and
    $c$-jets cannot be used for reconstruction efficiencies and smearing.}
  \label{tab:sfs_objs}
\end{table}

\begin{table}
  \centering
  \renewcommand{\arraystretch}{1.4}
  \setlength\tabcolsep{5pt}
  \begin{tabular}{c | c c}
    Properties & Symbol & Label \\  \hline
     $x$ momentum component & $p_x$               & \verb+PX+  \\
     $y$ momentum component & $p_y$               & \verb+PY+  \\
     $z$ momentum component & $p_z$               & \verb+PZ+  \\
     Transverse momentum & $p_T$               & \verb+PT+  \\
     Transverse energy   & $E_T$               & \verb+ET+  \\
     Energy              & $E$                 & \verb+E+  \\
     Pseudo-rapidity     & $\eta$ (or $|\eta|$)&\verb+ETA+ or (\verb+ABSETA+)\\
     Azimuthal angle (in $[0, 2\pi[$) & $\varphi$           & \verb+PHI+ \\
    \hdashline
     Number of tracks    & $n_{\rm tr}$ & \verb+NTRACKS+
  \end{tabular}
  \caption{\it Main object properties entering the different functions relevant
    for a simplified fast detector simulation in \madanalysis. The number of
    tracks $n_{\rm tr}$ can only be used for jet and hadronic tau (mis)tagging,
    when defining the domain of application of the corresponding efficiency
    piecewise function. The complete list of observables can be obtained from
    the manual~\cite{Conte:2012fm} or the
    normal mode reference card (see App.~A of Ref.~\cite{Conte:2018vmg}).}
  \label{tab:sfs_props}
\end{table}

The function representing the efficiency and its corresponding domain of
application (if relevant) have to be provided as valid formulas in {\sc Python}.
They can involve any observable supported by the \madanalysis\ interpreter, the
mostly relevant ones being the transverse momentum $p_T$ (\verb+PT+), the $x$,
$y$ and $z$ components of the momentum (\verb+PX+, \verb+PY+ and \verb+PZ+), the
pseudo-rapidity $\eta$ (\verb+ETA+) or its absolute value $|\eta|$
(\verb+ABSETA+), the energy $E$ (\verb+E+), the transverse energy $E_T$
(\verb+ET+) and the azimuthal angle $\varphi \in [0, 2\pi[$ (\verb+PHI+). All
these properties are collected in table~\ref{tab:sfs_props}.

For instance, in the following (toy) snippet of code, we define a flat photon
reconstruction efficiency of 99\% provided that the photon energy $E$ is larger
than 2~GeV and its pseudo-rapidity satisfies $|\eta|\leq0.88$. When the energy
$E>2$~GeV and $0.88<|\eta|<3$, the efficiency is of 98\%, and it finally
vanishes otherwise.
\begin{verbatim}
 define reco_efficiency a 0  \
    [E<=2. or ABSETA>=3.]
 define reco_efficiency a 0.99 \
    [E>2. and ABSETA<=0.88]
 define reco_efficiency a 0.98 \
    [E>2. and ABSETA>0.88 and ABSETA<3.]
\end{verbatim}
As can be seen, the usage of the \verb+or+ and \verb+and+ keywords is supported
(and needed) to define the different parts of the domain of definition of the
piecewise function.

\subsection{Smearing}\label{sec:smear}

The impact of the detector resolution is performed via the smearing formalism.
The particle momenta, as returned by the Monte Carlo simulations, are smeared by
\madanalysis\ on the basis of Gaussian functions of vanishing mean and a width
provided by the user.

The standard deviation of those Gaussian functions $\sigma$
directly depends on the object properties. In the following, we rely
on the example of energy smearing for calorimeter-based quantities. The
discussion can however be straightforwardly adapted to any other class of
energy-momentum smearing. In this example, energy smearing is often
parametrised as originating from three distinct components~\cite{Fabjan:2003aq},
\be\bsp
  \frac{\sigma(E,\eta,\varphi)}{E} = &\ \frac{N(\eta,\varphi)}{E} \oplus
    \frac{S(\eta,\varphi)}{\sqrt{E}} \oplus C(\eta,\varphi)\\
  = &\ \sqrt{ \frac{N^2(\eta,\varphi)}{E^2} + \frac{S^2(\eta,\varphi)}{E}+
     C^2(\eta,\varphi)} \ .
\esp\label{eq:sigpt}\ee
In this expression, $N$ (in GeV) corresponds to the so-called noise term
describing both the imperfections in the readout electronics, and the
fluctuations arising from the simultaneous energy deposits of uncorrelated
pile-up jets. This component of the resolution dominates for low-energy
objects. The second term, $S$ (in $\sqrt{\rm GeV}$), represents the stochastic
contribution related to statistical random fluctuations in the physical
evolution of the shower in the detector,
whereas the energy-indepen\-dent last term $C$ (therefore more relevant for
high-energy objects) is associated with imperfections in the calorimeter
geometry, anomalies in signal collection uniformities, as well as with
inter-calibration errors and fluctuations in the longitudinal energy content.

\begin{figure}
 \centering
 \includegraphics[scale=0.45]{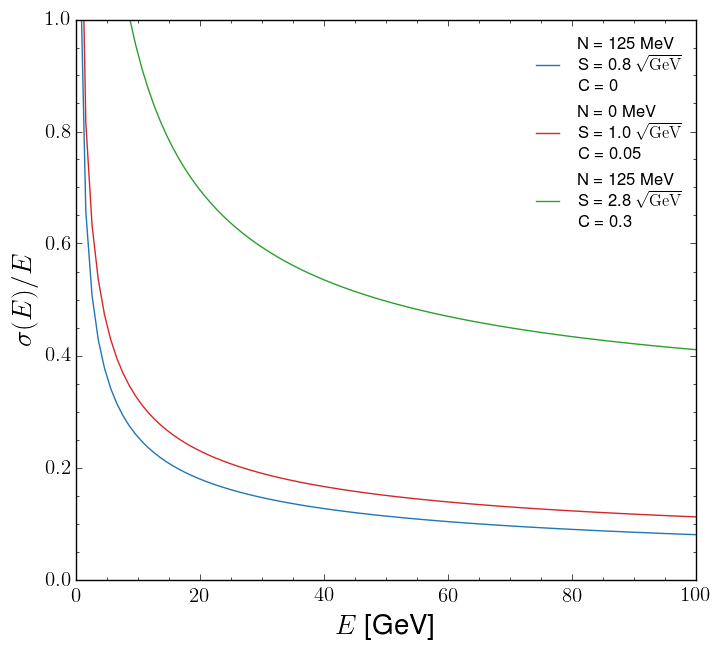}
 \caption{\it Energy resolution, given as a function of the energy, for a few
   configurations of the noise, stochastic and constant factors introduced in
   eq.~\eqref{eq:sigpt}.\label{fig:cal_res}}
\end{figure}

Several examples for the energy resolution dependence on the object energy are
shown in figure~\ref{fig:cal_res}, for three set of noise, stochastic and
constant
factor values. We first consider a case in which the resolution is essentially
dominated by its noise and stochastic terms (blue), so that it is especially
poor in the low-energy regime. We then focus on a setup on which the
noise factor is negligible with respect to a constant contribution (red),
resulting to a
quite similar overall resolution. Finally, a more realistic situation (green) in
which all three components are contributing is presented, the total
resolution being globally worse over the entire energy range.

In all those examples, the energy $E$ is smeared by a quantity distributed
according to a Gaussian of vanishing mean, and of standard deviation given by
eq.~\eqref{eq:sigpt},
\be
  E\to E^\prime=E + {\cal N}_\varepsilon\big[E; 0,\sigma(E,\eta,\varphi)\big]\ .
\label{eq:Esmear}\ee
The smeared energy $E'$ is in addition enforced to be positive and the quantity
${\cal N}_\varepsilon$ is a random parameter sampled according to the standard
normal distribution ${\cal N}$,
\be
  {\cal N}(x; \mu,\sigma) = \frac{1}{\sqrt{2\pi}\sigma}\ 
    \exp\bigg[-\frac12 \bigg(\frac{x-\mu}{\sigma}\bigg)^2\ \bigg]\ .
\ee

Smearing can be enabled in \madanalysis\ through the \verb+define+ command,
that takes as a first argument the \verb+smearer+ keyword indicating that a
smearer is being defined. The complete syntax reads
\begin{verbatim}
 define smearer <obj> with <obs> <func> [<dom>]
\end{verbatim}
The object whose properties have to be smeared is denoted by \verb+<obj>+ and
has to be provided either through its generalised PDG code, or its label (see
table~\ref{tab:sfs_objs}). The property that has to be smeared is provided
through the \verb+<obs>+ argument, the list of observables available being shown
in table~\ref{tab:sfs_props}. Finally, the function \verb+<func>+ (and its
corresponding domain of application \verb+<dom>+) represents the resolution
$\sigma$, that can be provided as a piecewise function depending on any of the
object properties.

After the smearing of some object properties, the missing
energy is always recalculated accordingly.

As an example, we present below a snippet of code that could be relevant for the
smearing of the jet energy at a typical LHC detector~\cite{%
Chatrchyan:2014fea},
\begin{verbatim}
 define smearer j with E \
   sqrt(E^2*0.05^2 + E*1.5^2) \
   [ABSETA <= 3]
 define smearer j with E \
   sqrt(E^2*0.13^2 + E*2.7^2) \
   [ABSETA > 3 and ABSETA <= 5]
\end{verbatim}
the piecewise function being only defined for cases in which the jets can
effectively be reconstructed. We indeed implicitly assume that the jet
reconstruction efficiency vanishes anywhere outside the considered domain. The
resolution $\sigma(E)$ corresponds to
\be
\renewcommand{\arraystretch}{1.4}
 \sigma(E) = \left\{\begin{array}{l l }
   0.05 \oplus 1.5 \sqrt{E} &\hspace{0.5cm}{\rm for~} |\eta|\leq 3 \ , \\
   0.13 \oplus 2.7 \sqrt{E} &\hspace{0.5cm}{\rm for~} 3<|\eta|\leq 5\ . \\
 \end{array}\right.
\label{eq:jet_ptsmearing}\ee

Whilst the methods that are presented above hold for any class of object, jet
smearing can be implemented in a second manner. Instead of smearing the
properties of each reconstructed jet, treating it as a whole, the code offers
the option to smear instead the properties of each constituent of the jet. The
resulting jet four-momentum is then evaluated in a second step. For most cases,
a jet-based smearing is sufficient to emulate most detector environments and
there is no need to rely on any constituent-based smearing. However, this by
definition renders the reconstruction process blind to any hadronic effect
occurring inside a jet. The latter may be relevant for very specific studies,
like when jet substructure is in order~\cite{Buckley:2019stt}. In
this case, hard QCD radiation can for instance activate detector cells around
those defining the jet and leads to the presence of two overlapping jets sharing
the energy deposits of the surrounding cells. Such an effect can be covered by
constituent-based jet smearing.

In order to activate constituent-based or jet-based jet smearing, it is
sufficient to type, in the \madanalysis\ command-line interface,
\begin{verbatim} set main.fastsim.jetrecomode = <val> \end{verbatim}
where \verb|<val>| can take either the \verb|jets| value (default), or the
\verb|constituents| value.

\subsection{Object identification - taggers}\label{sec:tag}

\begin{table}
  \centering
  \renewcommand{\arraystretch}{1.4}
  \setlength\tabcolsep{8pt}
  \begin{tabular}{c c}
    Reconstructed object & Truth object\\
    \hline
    $b$-jet & $b$-jet, $c$-jet, light jet\\
    $c$-jet & $b$-jet, $c$-jet, light jet\\
    Hadronic tau & Hadronic tau, light jet\\
    Photon & Light jet, electron, muon\\
    Electron & Light jet, muon, photon\\
    Muon & Electron, photon\\
  \end{tabular}
  \caption{\it List of tagging efficiencies and mistagging rates that are
    supported by the code.}\label{tab:tagger}
\end{table}

In its simplified detector simulation, \madanalysis\ allows the user to input
a large set of tagging efficiencies and related mistagging rates. The list of
available pairs of reconstructed and truth-level objects is shown in
table~\ref{tab:tagger}.

Jets can be (correctly and incorrectly) tagged as $b$-jet, $c$-jet or lighter
jets. Both light jets and true hadronic taus can be identified as taus. In
addition, jets faking electrons and photons,
as well as electrons, muons and photons faking each other can be implemented.
While the efficiencies relevant for $b$-tagging and tau-tagging can be
provided as well (those consist in `taggers' strictly speaking), the situation
is slightly different for electrons, muons and photons. Here, the tagging
efficiencies simply consist in the corresponding reconstruction efficiencies
(see section~\ref{sec:reco_eff}) and should be implemented as such. The
efficiency for correctly identifying a jet as a jet, a photon as a photon, an
electron as an electron and a muon as a muon cannot thus be implemented as
taggers, and the user has to enter them by means of the definition of
reconstruction efficiencies.

A (mis)tagger can be defined by typing, in the com\-mand-line interface,
\begin{verbatim} define tagger <true> as <reco> <func> [<dom>] \end{verbatim}
Once again, this relies on the \verb+define+ command, that takes as a first
argument the keyword \verb+tagger+ indicating to the code that a tagger is being
defined.
The keyword \verb+<true>+ represents the label of the truth-level object (see
table~\ref{tab:sfs_objs}) that can be tagged as a reconstructed object denoted
by \verb+<reco>+ (see table~\ref{tab:sfs_objs} as well). As in the previous
sections, the efficiency is provided as the function \verb|<func>|, to be
given as a valid {\sc Python} formula, and \verb|<dom>| consists in its domain
of application that is relevant when the (mis)tagging efficiency is a piecewise
function. For any tagger of a jet or a hadronic tau as any given object, it is
possible to make use of the number of tracks associated with the truth-level
object to define the domain of application of the different elements of the
piecewise function \verb+<func>+.

The definition of a tagger leads to the generation of a probability distribution
that indicates, at the time that the C++ code is executed, whether a
reconstructed object (a photon, a jet, \etc) or one of its properties
($b$-tag, $c$-tag, \etc) has to be modified.

As an illustrative example, $b$-tagging performances that are typical of an LHC
detector could be entered as
\begin{verbatim}
 define tagger b as b \
   tanh(PT/400)*(85/(4+0.252*PT)) \
   [ABSETA <= 2.5]
 define tagger b as b 0 [ABSETA > 2.5]

 define tagger c as b \
   tanh(0.018*PT)*(1/(4+0.0052*PT))

 define tagger j as b 0.01+3.8e-05*PT
\end{verbatim}
In this snippet of code, the $b$-tagging efficiency $\varepsilon_{b|b}$ is
parametrised by~\cite{Chatrchyan:2012jua}
\be
  \varepsilon_{b|b}(p_T) = 
    \tanh\big[p_T/400\big] \frac{85}{4+0.252 p_T}\Theta\big[2.5-|\eta|\big]\ ,
\label{eq:btag1}\ee
whilst the mistagging rate of a charmed jet ($\varepsilon_{b|c}$) as a $b$-jet
and the one of a light jet as a $b$-jet ($\varepsilon_{b|j}$) are given by
\be\bsp
  \varepsilon_{b|c}(p_T) = &\
    \tanh\big[0.018 p_T\big] \frac{1}{4+0.0052 p_T}\ ,\\
  \varepsilon_{b|j}(p_T) = &\
    0.01  + 3.8\!\cdot\! 10^{-5} p_T\ .
\esp\label{eq:btag2}\ee

\subsection{Expert mode}\label{sec:expert}

Whereas the metalanguage defining the \madanalysis\ interpreter is rich, it is
limited by its own definition. The user can circumvent this inherent limitation
by using the platform in its so-called expert mode. Analyses are here
implemented directly in C++, bypassing the \madanalysis\ command-line interface.
In this way, the user
can benefit from all capabilities of the programme (readers, observables, \etc),
and focus on implementing only the non-standard routines necessary for his/her
own purpose.

The analysis skeleton generated automatically when the expert mode is switched
on can incorporate a simplified detector simulation as defined in the beginning
of this section. To this aim, it is sufficient to initiate \madanalysis\ by
typing in a shell,
\begin{verbatim}
 ./bin/ma5 -Re <wdir> <label> <sfs>
\end{verbatim}
where \verb|<wdir>| stands for the working directory in which the analysis
template is generated, \verb|<label>| refers to the analysis name that is used
throughout the entire analysis, and \verb|<sfs>| is the (optional)
configuration file,
written as a set of normal \madanalysis\ commands, defining how the simplified
fast detector simulation should be run. Providing such a file is optional, so
that if it is absent, the code runs as in previous versions of the programme. We
refer to the
manual~\cite{Conte:2014zja} of the expert mode and ref.~\cite{Conte:2018vmg} for
more information.

\subsection{Standard LHC detector parametrisation}\label{sec:lhc}
\madanalysis\ is shipped with two predefined detector parametrisations, that
are respectively related to the ATLAS and CMS detectors. Those cards are
available from the \verb+madanalysis/input+ subdirectory, and have been
validated through a comparison with the standard ATLAS and CMS \delphes\
detector configuration files. The example of the CMS validation is presented
in section~\ref{sec:validation}.

In order to load those cards when running \madanalysis, the user has to start
the code by providing the path to the detector card as an argument of the
\verb+bin/ma5+ command. This would give, from a shell,
\begin{verbatim}
 ./bin/ma5 -R madanalysis/input/<card>.ma5
\end{verbatim}
where \verb+<card>+ has to be replaced by \verb+CMS_default+ and
\verb+ATLAS_default+ for the CMS and ATLAS detector pa\-ra\-me\-tri\-sa\-ti\-ons
respectively.

\section{Comparison with \delphes\ and the Monte Carlo truth}
\label{sec:validation}

In this section, we validate the implementation of our simplified fast detector
simulation in \madanalysis. To this aim, we perform a comparison between
predictions relying on our module and predictions relying instead on \delphes,
for a variety of Standard Model processes. Both results are moreover confronted
to the expectation of the Monte-Carlo truth, where events are reconstructed as
such (without any smearing and tagging). The general design of our simulation
framework is depicted in section~\ref{sec:design_val}, whilst
sections~\ref{sec:QCD}, \ref{sec:DY}, \ref{sec:tau} and \ref{sec:photons} focus
on jet, lepton, hadronic tau and photon properties respectively.

\subsection{Simulation framework}\label{sec:design_val}

In order to validate our implementation, we generate several samples of Monte
Carlo events describing various Standard Model processes relevant for
proton-proton collisions at a centre-of-mass energy of 14~TeV. Hard scattering
events are produced with \madgraph~\cite{Alwall:2014hca}
(version 2.7.3), that we use to
convolute leading-order matrix elements with the leading-order set of NNPDF~2.3
parton densities~\cite{Ball:2012cx}. Those events are matched with parton
showering as modelled by the \pythia\ package~\cite{Sjostrand:2014zea}
(version 8.244), that is
also employed to simulate hadronisation. As we focus on Standard Model
processes, the typical energy scales under consideration are not so hard,
ranging between a few tens of GeV to 100--200 GeV. Section~\ref{sec:pad} will be
instead dedicated to new physics, and will thus be relevant for a
study of the features of our fast detector simulation framework for much larger
scales.

In our comparison, we include detector effects in four different ways. First, we
consider an
ideal detector, or the so-called Monte Carlo truth. In practice, reconstruction
is performed as described in section~\ref{sec:reco}, with all detector effects
being switched off. We rely on the anti-$k_T$ algorithm with a radius parameter
set to $R=0.4$, as implemented in \fastjet\ (version 3.3.3).
Second, we make use of the complex detector machinery as
implemented in the \delphes\ package (version 3.4.2),
using the standard CMS detector
parametrisation shipped with the programme. In particular, this
includes object isolation
and enforces the storage of all generator-level
objects (as for the SFS for what concerns the hard-scattering process) and
calorimetric and tracking information. Finally, we consider the simplified
fast detector simulation presented in this work, that we run once with a
jet-based jet smearing and once with a constituent-based jet smearing. The
detector parametrisation has been designed such that it matches the standard CMS
card from \delphes.

By definition, object isolation cannot be
implemented directly within the SFS framework, so that it needs
to be performed at the analysis level. In this paper, we provide one example of
a way to handle this, which relies on a simple method employing
$\Delta R$-based object separation.
We consider two objects as overlapping if they lie at a
distance $\Delta R$ in the transverse plane that is smaller than some threshold.
An overlap removal procedure is then implemented, one of the two objects being
removed. The details about which object is kept and which object is removed
depend on what the user has in mind. We recall that in the entire SFS framework,
those objects always refer to objects that are not originating from any hadronic
decay processes, those latter objects being instead clustered into jets.

In sections~\ref{sec:QCD},
\ref{sec:tau} and \ref{sec:photons}, we focus on jets, taus and photons.
We therefore first remove from the jet collection any jet that would be too
close to an electron. We then remove from the electron, muon and photon
collections any electron, muon and photon that would be too close to any of the
remaining jets. Finally, taus are required to be well separated from any jet.
This strategy is similar to what is done,
{\it e.g.} in the ATLAS analysis of ref.~\cite{Aad:2019pfy}, that examines the
properties of hadronic objects. In section~\ref{sec:DY}, we focus instead on
leptons and therefore implement a different isolation methodology inspired by
the ATLAS analysis of ref.~\cite{Aad:2019vnb}. Here, we
firstly remove any lepton that would be too close to a jet, before secondly
removing any jet that would be too close to any of the remaining leptons.
More information about object isolation is provided, specifically for the
different cases under study, in sections~\ref{sec:QCD}, \ref{sec:DY},
\ref{sec:tau} and \ref{sec:photons}.
It is important to bare in mind that this crude approximation (relatively to the
\delphes\ capabilities) for isolation is sufficient
in many useful physics cases, as testified by the results shown in
section~\ref{sec:pad} in the
context of LHC recasting. Depending on what the user aims to do, this may
however be insufficient.

\begin{table}
  \centering
  \renewcommand{\arraystretch}{1.4}
  \setlength\tabcolsep{5.5pt}
  \begin{tabular}{c|c c c}
    $N_{\rm events}$ & \delphes & SFS [Jets] & SFS [Const.] \\\hline
    10,000  & 821.0 Mb (8.7~Mb)& 6.7  Mb & 6.7 Mb\\
    50,000  & 4.1 Gb   (43~Mb)& 34.0 Mb & 34.0 Mb\\
    100,000 & 8.1 Gb   (84~Mb)& 67.0 Mb & 68.0 Mb\\
    200,000 & 17 Gb    (166~Mb)& 133 Mb  & 134 Mb\\[-.3cm]
    \multicolumn{4}{c}{}\\[.02cm]
    $N_{\rm events}$ & \delphes & SFS [Jets] & SFS [Const.] \\\hline
    10,000  & 3 min 24 s  & 2 min 11 s  & 2 min 17 s \\
    50,000  & 17 min 15 s & 10 min 24 s & 11 min 02 s \\
    100,000 & 33 min 21 s & 20 min 41 s & 22 min 20 s \\
    200,000 & 1 h 07 min  & 42 min 07 s & 43 min 55 s
  \end{tabular}
  \caption{\it Comparison of the reconstruction of $N_{\rm events}$ top-antitop
    events with \delphes, using the built-in CMS detector
    parametrisation, and the simplified fast detector simulation introduced
    in this work. We consider jet smearing based either on the reconstructed
    jets themselves (SFS [Jets]) or on their constituents (SFS[Const.]). The
    results depict the size of the output file when the results are
    compressed (top panel) and the time needed
    for the run (lower panel), as obtained by using a machine with an Intel(R)
    Xeon(R) Core E3-1271v3 CPU with a 3.6 GHz clock speed. The
    time necessary for the input/output operations is negligible, so that the
    provided results can be considered as related to the detector simulation
    process only. In the upper
    table, we also provide information, included in parentheses, on the output
    size corresponding to a \delphes\ parametrisation in which the calorimetric,
    track, jet substructure and generator-level information is not written to
    the file.
    \label{tab:speed-space} }
\end{table}

As mentioned in the introduction to this paper, our new module has the
advantage on \delphes\ to be cheaper in terms of CPU costs
when its default configuration is compared with the default ones
implemented in \delphes\ (that many users employ).
This is illustrated in table~\ref{tab:speed-space} for the
reconstruction of different numbers of top-antitop events $N _{\rm events}$.
Such a process leads to hadron-level events including each about
1670 objects, that thus consist of the amount of inputs to be processed at the
time of the simulation of the detector effects. It is interesting
to note that in the
context of LHC recasting (as discussed in section~\ref{sec:pad}), the speed
difference often reaches a factor of 2. This extra gain originates from the
structure of the PAD itself. When relying on the SFS framework for the
simulation of the detector, the analysis and the detector simulation can be
performed directly, without having to rely on writing the reconstructed events
on a file. In contrast, when relying on \delphes\ for the simulation of the
detector, an intermediate \rooot\ file is first generated by \delphes\ and then
read again at the time of the analysis. This results in a loss of efficiency in
terms of speed.

It can also be seen, in table~\ref{tab:speed-space},
that regardless of the size of the event sample, our simplified
fast detector simulation is up to 30\% faster  than \delphes.
As the time necessary for the input/output operations is negligible,
the numbers provided in the lower panel of the table can be seen as what is
needed to deal with the simulation of the detector itself. Independently of the
event sample size, \delphes\ is found to process one event in about 0.0203~s,
whilst the SFS framework requires 0.0130~s instead, in average and regardless of
the jet smearing configuration. In addition, we have compared the SFS
performance with those of \delphes\ when relying on a much simpler detector
parametrisation that only includes elementary smearing, object reconstruction
and identification functionalities. Isolation and tracking are thus turned off,
so that we make use of \delphes\ in a way that is as close as possible as what
the SFS does. We observe time performances that are slightly closer to the SFS
ones, \delphes\ requiring here only 0.0189~s to process a single event in
average.

In terms of the disk space needed to store the outputed
reconstructed events, we have found that \delphes, when used in its default
configuration, leads to output files that are about 100 times heavier than for
an SFS-based detector simulation. However,
most of the disk space is used to store
track and calorimetric tower information. As this information is not included in
the SFS framework, we investigate how the file sizes change when it is removed
from the default CMS parametrisation in \delphes. We also remove from the output
file any information related to the generator-level collections and jet
substructure. We consequenty obtain results similar
in the SFS and \delphes\ cases, as visible from the upper panel of
table~\ref{tab:speed-space}.

\subsection{Multijet production}\label{sec:QCD}

\begin{figure*}
  \centering
  \includegraphics[scale=0.44]{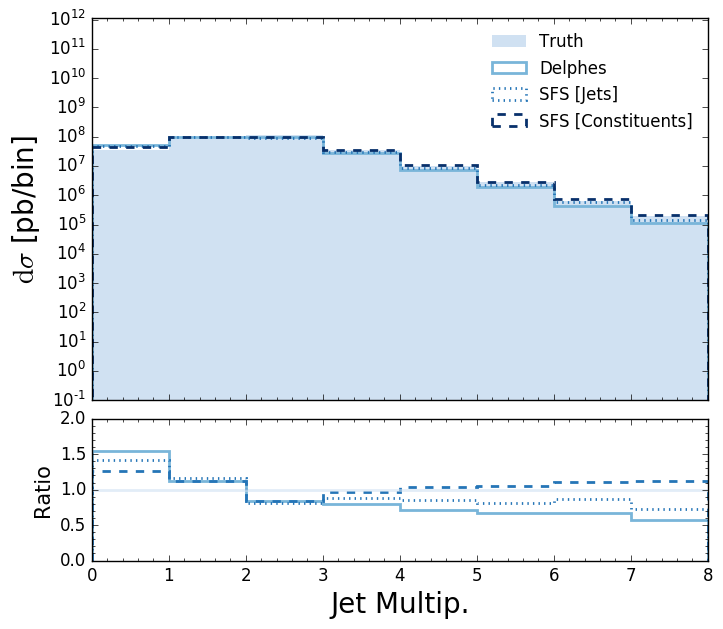}
  \includegraphics[scale=0.44]{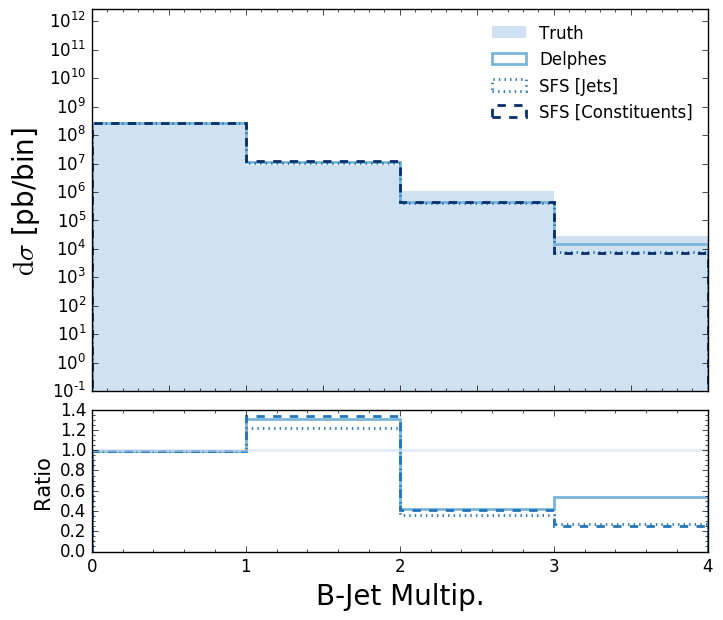}\\
  \includegraphics[scale=0.44]{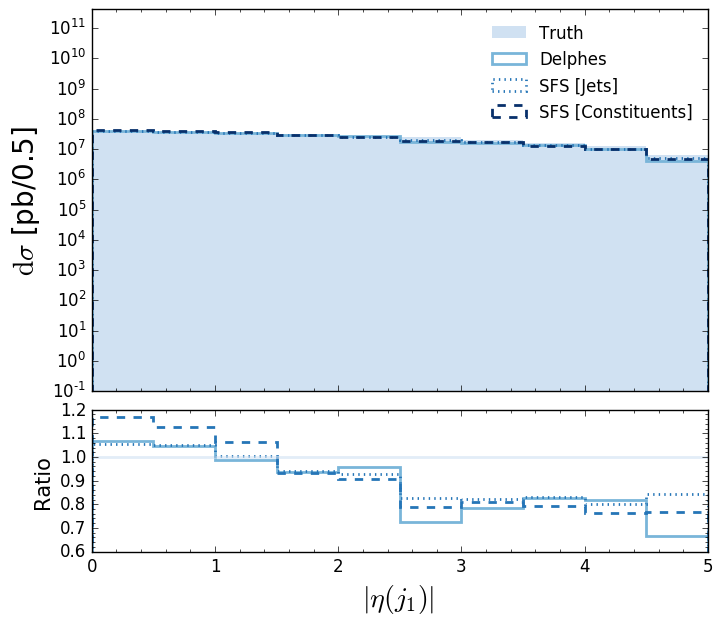}
  \includegraphics[scale=0.44]{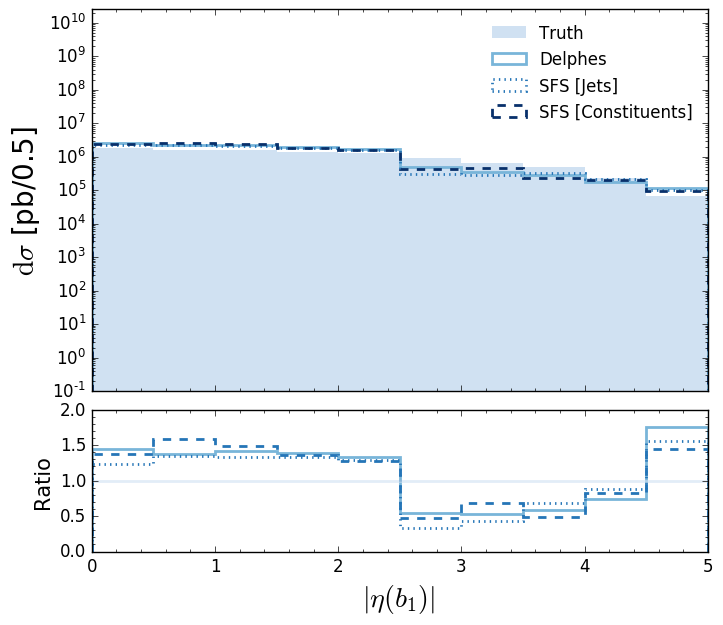}\\
  \includegraphics[scale=0.44]{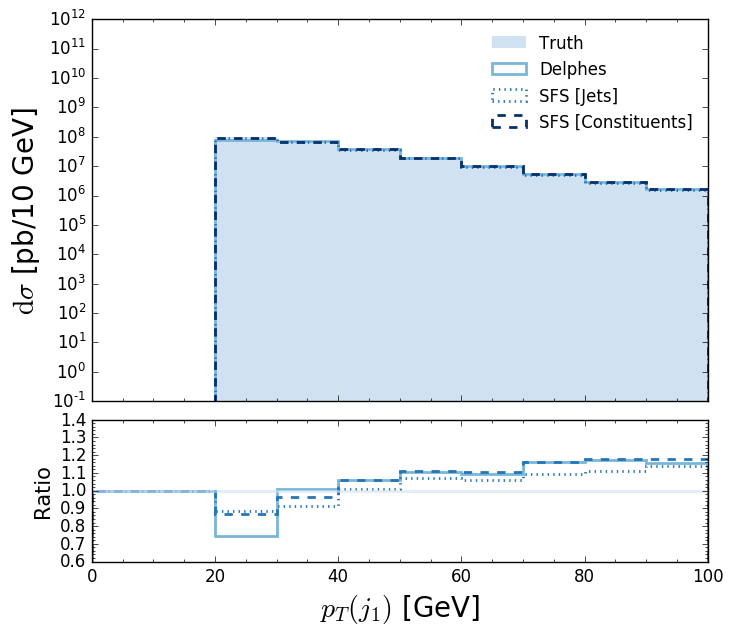}
  \includegraphics[scale=0.44]{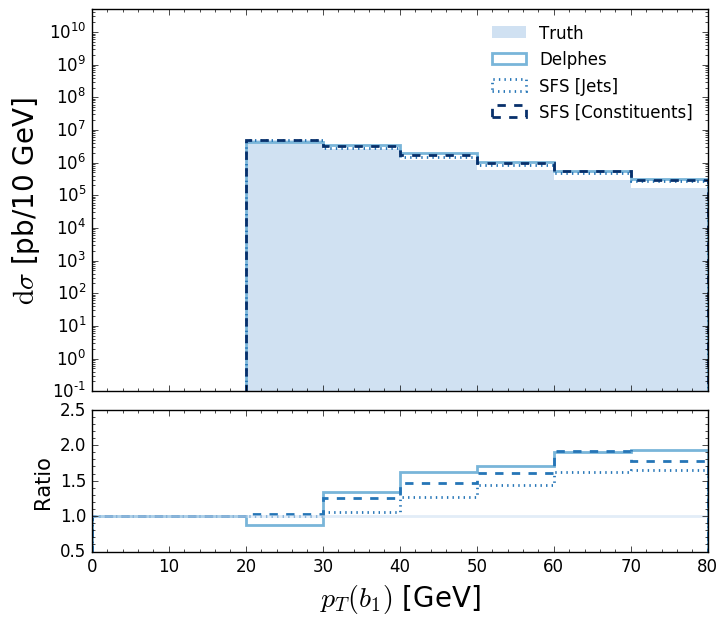}
\caption{\it Jet properties when reconstruction is considered with various
  options for the detector simulation: an ideal detector (truth, filled area),
  \delphes\ (solid) and the \madanalysis\ simplified fast detector simulation
  with either jet-based (SFS [Jets], dotted) or constituent-based (SFS
  [Constituents], dashed)
  jet smearing. We show distributions in the jet (upper left) and $b$-jet
  (upper right) multiplicity, leading light jet $p_T$ (centre left) and $|\eta|$
  (centre right), as well as leading $b$-jet $p_T$ (lower left) and $|\eta|$
  (lower right). In the lower insets, the distributions are normalised to the
  truth results.\label{fig:jet_props}}
\end{figure*}

In order to investigate the differences in the jet properties that would arise
from using different reconstruction methods, we consider hard-scattering di-jet
production, $pp\to j j$, in a five-flavour-number scheme. In
our simulation process, we generate
500,000 events and impose, at the generator
level, that each jet features a transverse momentum of at least 20~GeV and a
pseudo-rapidity smaller than 5. Moreover, the invariant mass of the di-jet
system is enforced to be larger than 100~GeV.

In our simplified fast detector simulation (SFS), the energy $E$ of
the jets is smeared
according to the transfer functions provided in ref.~\cite{Chatrchyan:2014fea}.
The resolution on the jet transverse momentum $\sigma(E)$ is given
by eq.~\eqref{eq:jet_ptsmearing}.
Moreover, we implement the $b$-tagging performance defined by
eqs.~\eqref{eq:btag1} and \eqref{eq:btag2}, keeping in mind that $b$-jet
identification is ineffective for $|\eta|>2.5$, by virtue of the CMS tracker
geometry. Finally, we include the following jet reconstruction efficiency
$\varepsilon_j(\eta)$, that depends on the jet pseu\-do-rapidity,
\be\renewcommand{\arraystretch}{1.4}
  \varepsilon_{j}(\eta) = \left\{\begin{array}{l l }
  92.5\%  & \hspace{0.5cm}{\rm for~}  |\eta| \leq 1.5\ ,\\
  87.5\%  & \hspace{0.5cm}{\rm for~}  1.5 < |\eta| \leq 2.5\ , \\
  80\%    & \hspace{0.5cm}{\rm for~}  |\eta| > 2.5 \ .
  \end{array}\right.
\label{eq:tracker_j}\ee
This allows for the simulation of most tracker effects on the jets, with the
limitation that charged and neutral jet components are equally considered.
Our detector simulation also impacts all the other objects potentially
present in the final state. However, we refer to the next sections for details
on the simulation of the corresponding detector effects.

At the analysis level, we select as jet and lepton candidates those jets and
leptons with a transverse momentum greater than 20~GeV and 10~GeV respectively.
We moreover impose simple isolation requirements by ignoring any
jet lying at an angular distance smaller than 0.2 of any reconstructed
electron ($\Delta R_{ej} < 0.2$), and then ignore any
lepton (electron or muon) lying at a distance smaller than 0.4 of
any of the remaining
jets ($\Delta R_{\ell j} < 0.4$).

In figure~\ref{fig:jet_props}, we present and compare various distributions when
an ideal detector is considered (filled areas, truth), when a CMS-like detector
is considered and modelled in \delphes\ (solid) and when the simplified fast
detector
simulation (SFS) introduced in this work, parametrised to model a CMS-like
detector and run both when jet-based (dotted) and constituent-based (dashed) jet
smearing is switched on. Consequently to the simple jet smearing
studied in this section, there is no need to implement different efficiencies
for constituent-based and jet-based smearing. However, this is not the case
anymore for investigations relying, for example, on the jet spatial
resolution like when jet substructure is involved.

\begin{figure}
  \centering
  \includegraphics[scale=0.46]{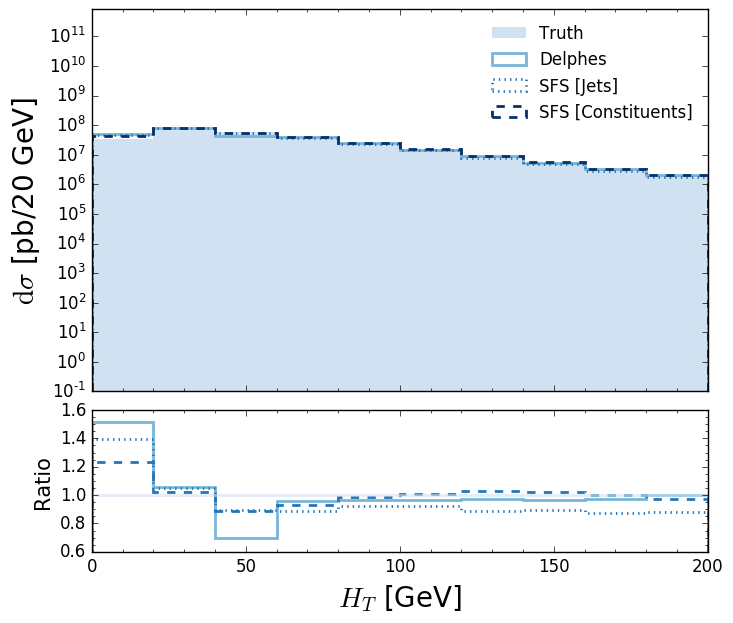}
  \caption{\it Same as figure~\ref{fig:jet_props}, but for the $H_T$
    spectrum.\label{fig:jet_HT}}
\end{figure}

In the upper line of the figure, we begin with studying the distribution in the
number of jets $N_j$ (upper left) and $b$-jets (upper right). As can be noticed
by inspecting the $N_j$ spectrum, the impact of the detector is especially
important for
the 0-jet bin, as well as when the number of jets is large. In this case,
deviations from the Monte Carlo truth are the largest, regardless of the way the
detector effects are simulated. However, \delphes-based or SFS-based
predictions agree quite well, at least when a jet-based smearing is
used in the SFS case. A constituent-based smearing indeed yields quite
significant differences, and predictions for the large-multiplicity bins are
much closer to the Monte Carlo truth. The detector imperfections
yield event migration from the higher multiplicity to the lower
multiplicity bins independently of the exact details those imperfections are
dealt with at the simulation level.

\begin{figure*}
  \centering
  \includegraphics[scale=0.44]{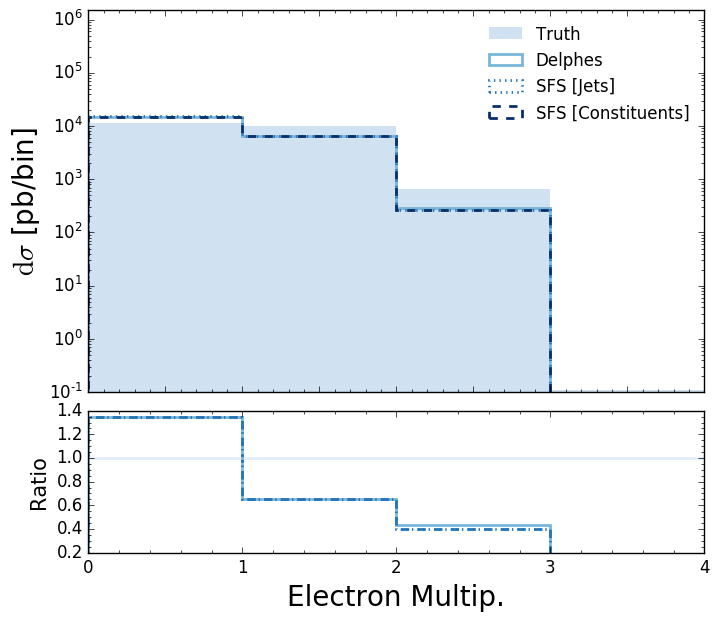}
  \includegraphics[scale=0.44]{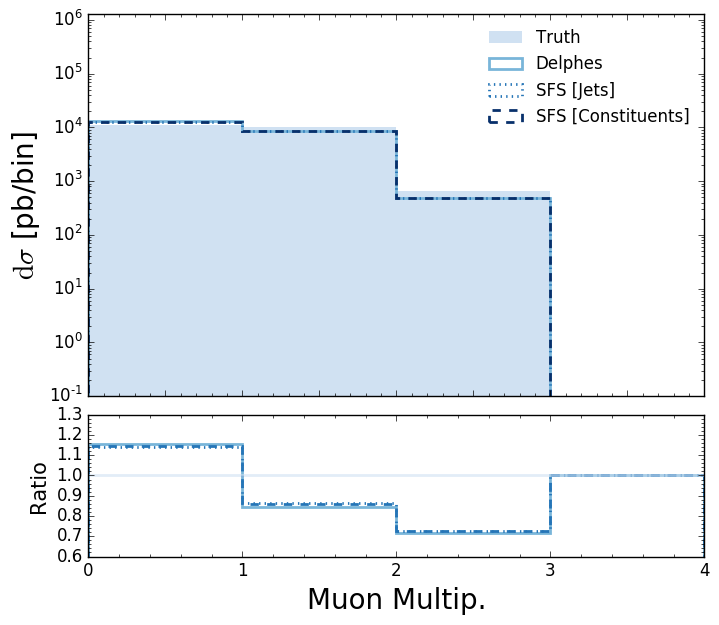}\\
  \includegraphics[scale=0.44]{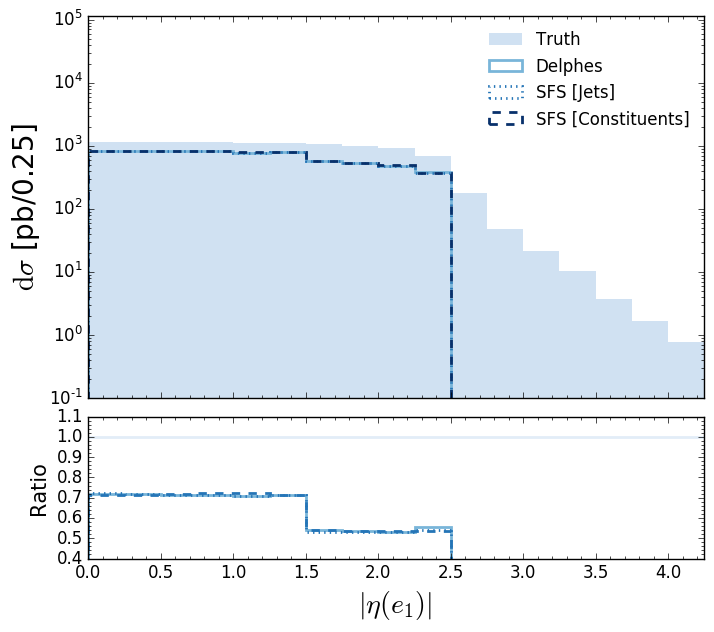}
  \includegraphics[scale=0.44]{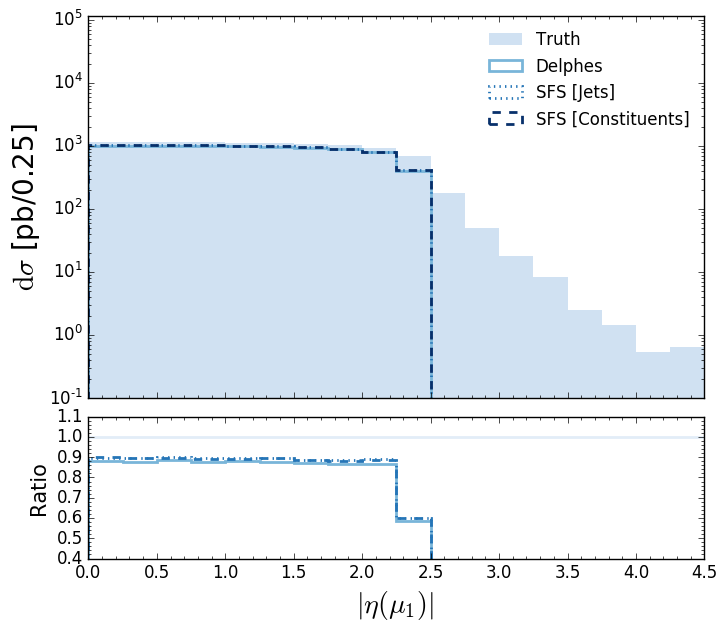}\\
  \includegraphics[scale=0.44]{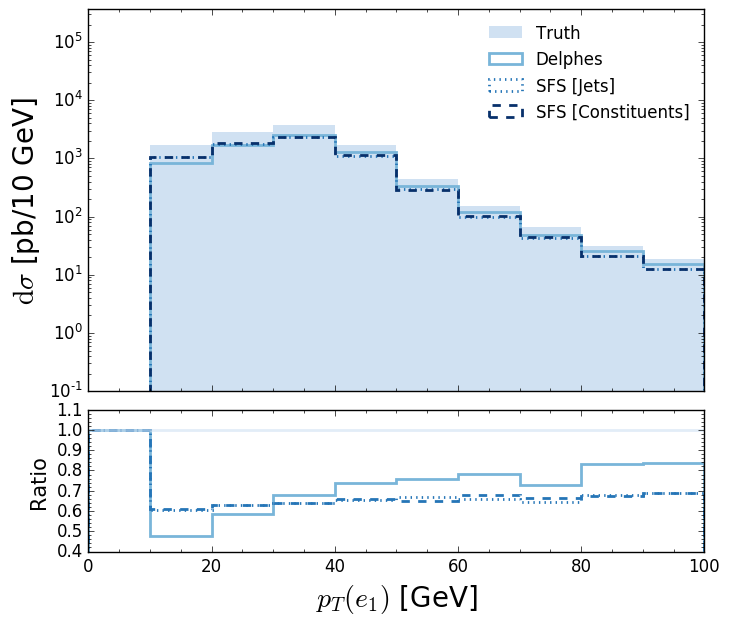}
  \includegraphics[scale=0.44]{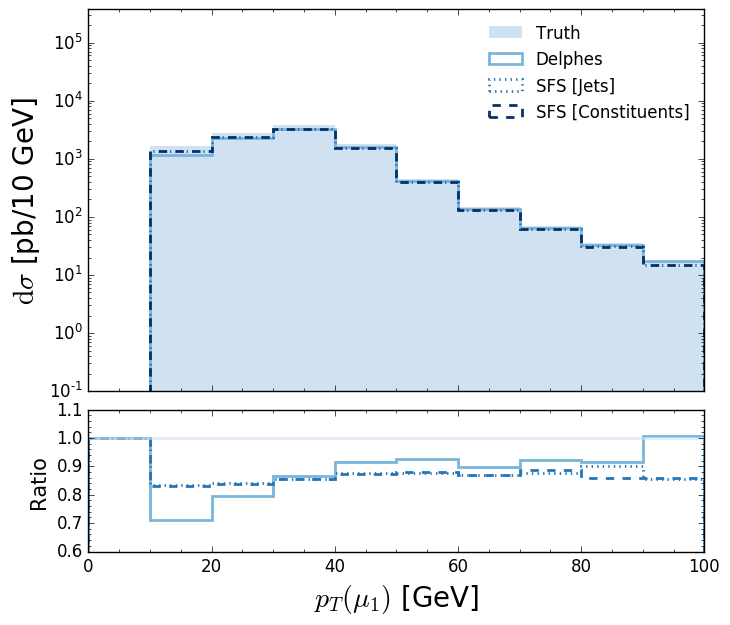}
\caption{\it Same as figure~\ref{fig:jet_props}, but for the electron and
  muon multiplicity distribution (upper row), and the pseudo-rapidity (central
  row) and transverse momentum (lower row) spectrum of the leading electron
  (left) and muon (right). \label{fig:lep_props}}
\end{figure*}

With the exception of the jet multiplicity spectrum,
the impact of the used jet smearing method is however
quite mild compared to the bulk of the detector effects. This is expected, as
those effects are important only in very specific cases (not covered in this
work). In the rest of the figure, we demonstrate this by investigating the
shape of various differential spectra, and comparing them with predictions
derived from the Monte Carlo truth. We study the transverse momentum and
pseudo-rapidity
spectrum of the leading jet and leading $b$-jet. All detector simulations agree
with each other. The existing differences between \delphes\ and the other
fast simulation results are found to originate from the reconstruction
efficiency (see
eq.~\eqref{eq:tracker_j}). The latter, as implemented in the simplified fast
detector simulation of \madanalysis, are expected to mimic, but only to some
extent, the tracker effects included in \delphes\ that are much more complex and
distinguish charged and neutral hadrons.
Those difference however only impact the soft parts of the spectrum.

Finally, we
consider a more inclusive variable in figure~\ref{fig:jet_HT}, namely the
scalar sum of the transverse momentum of all reconstructed jets $H_T$. We again
get a satisfactory level of agreement
between the three detector simulations. The
differences between \delphes-based and SFS-based predictions only
affect the low-$H_T$ bins in which the impact of the softest objects, that are
most likely to be sensitive to the different treatment of the detector
simulation, is the largest.

\subsection{Electrons and muons}\label{sec:DY}

In this section, we perform on a comparison of electron and muon properties when
these are reconstructed from the different methods considered in this work.
To this aim, we generate a sample of 200,000 neutral-current and
charged-current Drell-Yan events,
$pp\to\ell^+\ell^- + \ell^-\bar\nu_\ell + \ell^+\nu_\ell$ with
$\ell=e$, $\mu$. In our simulations, we impose, at the Monte Carlo generator
level, the invariant mass of a same-flavour opposite-sign lepton
pair to be of at least 50~GeV, and we constrain each
individual lepton to feature a transverse momentum greater than 10~GeV.

The electron and muon reconstruction efficiencies
$\varepsilon_e$ and $\varepsilon_\mu$ are extracted from
ref.~\cite{Khachatryan:2015hwa}, and depend both on the electron and muon
transverse momentum $p_T$ and pseudo-rapidity $\eta$. They respectively read
\be{\small\renewcommand{\arraystretch}{1.4}\bsp
  \varepsilon_e(p_T, \eta) \!=\!&
  \left\{\begin{array}{l l}
    0.7   & \hspace{0.15cm}{\rm for~}
       p_T\!>\!10~{\rm GeV~and~} |\eta|\!\leq\! 1.5\ ,\\
    0.525  & \hspace{0.15cm}{\rm for~}
       p_T\!>\!10~{\rm GeV~and~} 1.5\!<\! |\eta| \!\leq\! 2.5\ ,\\
    0     & \hspace{0.15cm}{\rm otherwise}\ ,
  \end{array}\right.\\
  \varepsilon_{\mu}(p_T, \eta) \!=\!&
  \left\{\begin{array}{l l}
  0.891\ \mathcal{B}(p_T) & \hspace{0.08cm}{\rm for~}
     p_T\!>\!\!10 {\rm GeV~and~} |\eta|\!\leq\! 1.5\ ,\\
  0.882\ \mathcal{B}(p_T) & \hspace{0.08cm}{\rm for~}
     p_T\!>\!\!10 {\rm GeV~and~} 1.5\!<\!|\eta|\!\leq\! 2.4\ ,\\
    0      & \hspace{0.08cm}{\rm otherwise}\ ,
  \end{array}\right.
\esp}\ee
with
\be
  \mathcal{B}(p_T) = \Theta\bigg[1 - \frac{p_T}{{\rm TeV}}\bigg] - 
     e^{\big[\frac12(1- p_T/{\rm TeV})\big]}
     \Theta\bigg[\frac{p_T}{\rm TeV}-1\bigg]\ .
\ee
In addition, we model the effects stemming from the electromagnetic
calorimeter by smearing the electron energy in a Gaussian way, with a standard
deviation $\sigma_e(E,\eta)$ defined by~\cite{Chatrchyan:2013dga,
Khachatryan:2015hwa}
\be\renewcommand{\arraystretch}{1.4}
  \frac{\sigma_e(E,\eta)}{E} = \left\{\begin{array}{l}
    \Big[1+0.64 |\eta|^2\Big] \
      \Big[\frac{0.4}{E}\oplus\frac{0.11}{\sqrt{E}}\oplus0.008\Big]\\
      \hspace{1.2cm}{\rm for~} |\eta| \leq 1.5 \\
    \Big[2.16 + 5.6(|\eta|\!-\!2)^2\Big] 
      \Big[\frac{0.4}{E}\oplus\frac{0.11}{\sqrt{E}}\oplus0.008\Big]\\
      \hspace{1.2cm}{\rm for~} 1.5 \!<\!|\eta| \!\leq\! 2.5\ , \\
     \frac{ 2.08}{\sqrt{E}}\oplus0.107\\
      \hspace{1.2cm}{\rm for~} 2.5 \!<\!|\eta| \!\leq\! 5\ . \\
  \end{array}\right.
\label{eq:e_smear}\ee

At the analysis level, we implement similar selections as in
section~\ref{sec:QCD}. We consider as jets and leptons those jets and leptons
with $p_T > 20$~GeV and 10~GeV respectively, and require all studied objects to
be isolated. Any lepton too close to a jet is discarded ($\Delta R_{\ell j}<
0.2$), and any jet too close to any of the remaining leptons is discarded too
($\Delta R_{\ell j} < 0.4$). Moreover, we impose a selection on the invariant
mass of the lepton pair, $m_{\ell\ell}>55$~GeV, for events featuring at least
two reconstructed leptons of opposite electric charges.

In figure~\ref{fig:lep_props}, we compare various lepton-related distributions
after using the different considered options for the simulation of the detector
effects. We
start by considering the distributions in the electron (upper left)
and muon (upper right) multiplicity, which show that the loss in leptons
relatively to the Monte-Carlo truth is similar for all three detector simulations.
This is not surprising as the efficiencies have been tuned accordingly.

For similar reasons, the detector effects impacting the leading electron (centre
left) and muon (centre right) pseu\-do-rapidity distributions are similarly
handled in all three setups. Relatively to the Monte-Carlo truth, not a single
electron and muon is reconstructed for pseudo-rapidities larger than 2.5 and 2.4
respectively, and the fraction of lost leptons is larger for $|\eta|>1.5$ than
for $|\eta|<1.5$. This directly stems from the detector geometry
and design, that make it impossible to reconstruct any non-central lepton and
lead to degraded performance for larger pseudo-rapidities (as implemented in all
detector simulator reconstruction efficiencies).

The various options for modelling the detector effects however yield important
differences for the leading electron (lower left) and leading muon (lower right)
transverse momentum distributions. \delphes\ predicts a smaller
number of leptons featuring $p_T<40$~GeV than in the SFS case, and a larger
number of leptons with transverse momenta $p_T > 40$~GeV.
The difference ranges up to 20\%
at the kinematical selection threshold of $p_T\!\sim\!10$~GeV,
and for the largest $p_T$ bins. This originates
from the inner machineries implemented in the various codes. In \delphes,
hadron-level objects are first converted into tracks and calorimetric deposits,
which involves efficiencies provided by the user. Next, momenta and
energies are smeared, before that the code reconstructs the objects to be used
at the analysis level. This latter step relies again on user-defined
reconstruction efficiencies, this time specific to each class of reconstructed
objects. In the \madanalysis\ SFS, such a splitting of the reconstruction
efficiency into two components has not been implemented, for the purpose of
keeping the detector modelling simple. Such a two-component smearing
would however introduce significant shifts in the distributions describing the
properties of the reconstructed objects,
especially for objects for which tracking effects matter.

\begin{figure}
  \centering
  \includegraphics[scale=0.46]{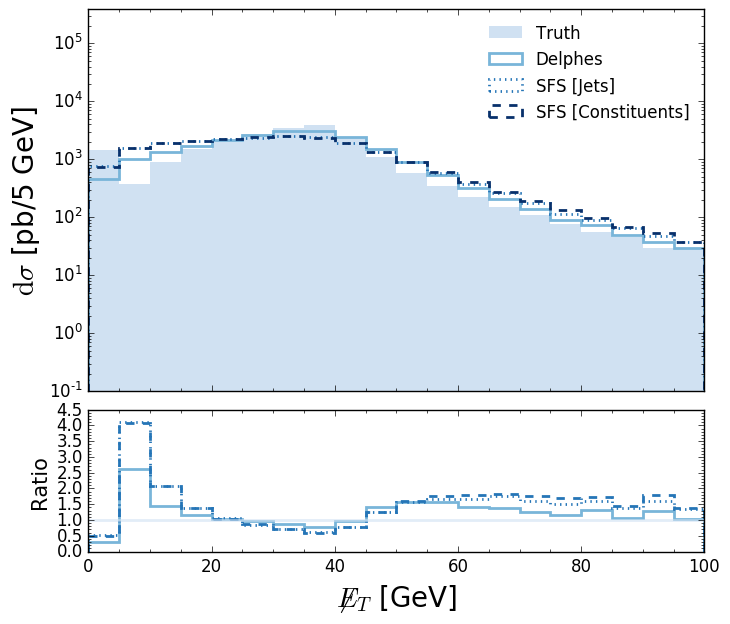}
  \caption{\it Same as figure~\ref{fig:jet_props}, but for the missing
    transverse energy spectrum.\label{fig:dy_met}}
\end{figure}

\begin{figure*}
  \centering
  \includegraphics[scale=0.44]{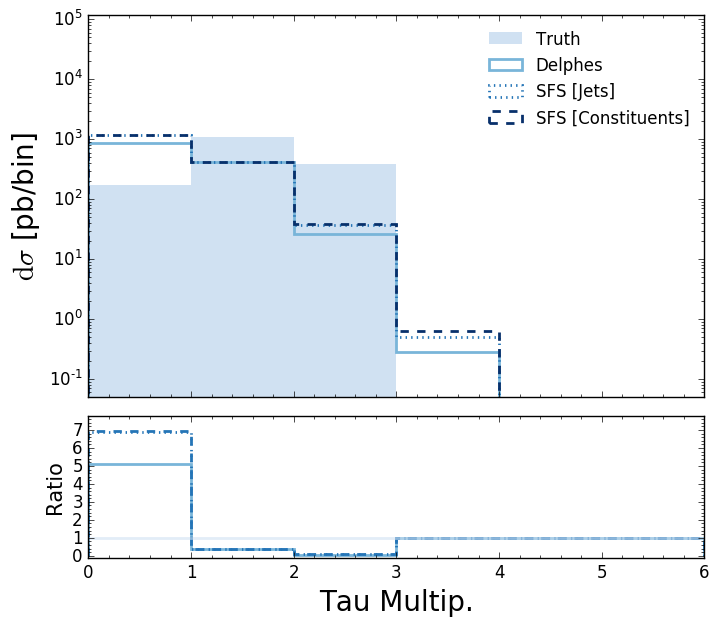}
  \includegraphics[scale=0.44]{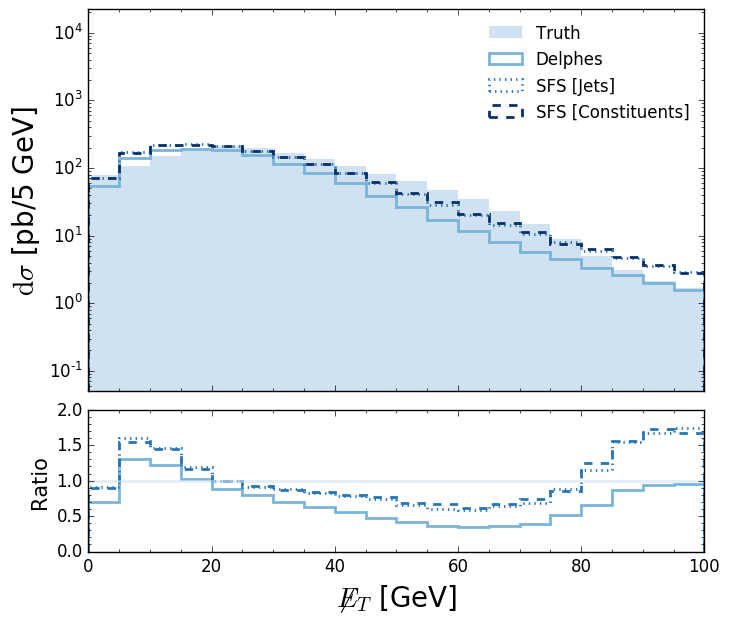}\\
  \includegraphics[scale=0.44]{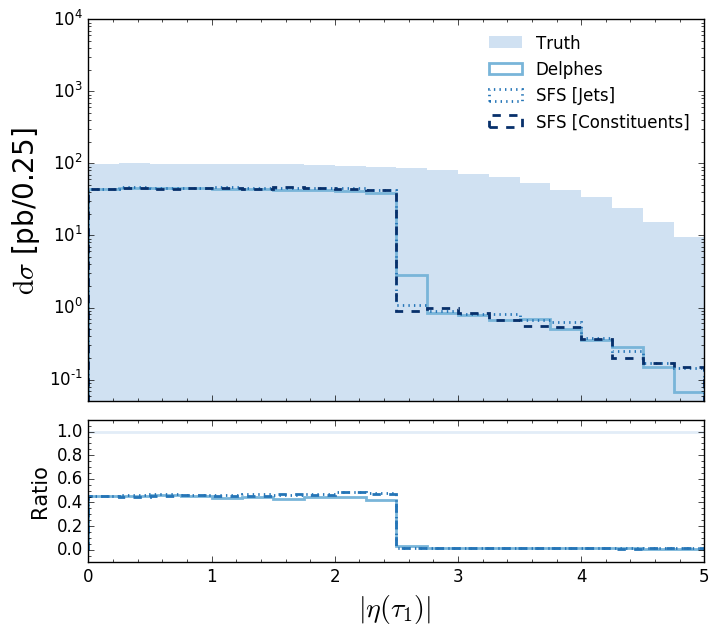}
  \includegraphics[scale=0.44]{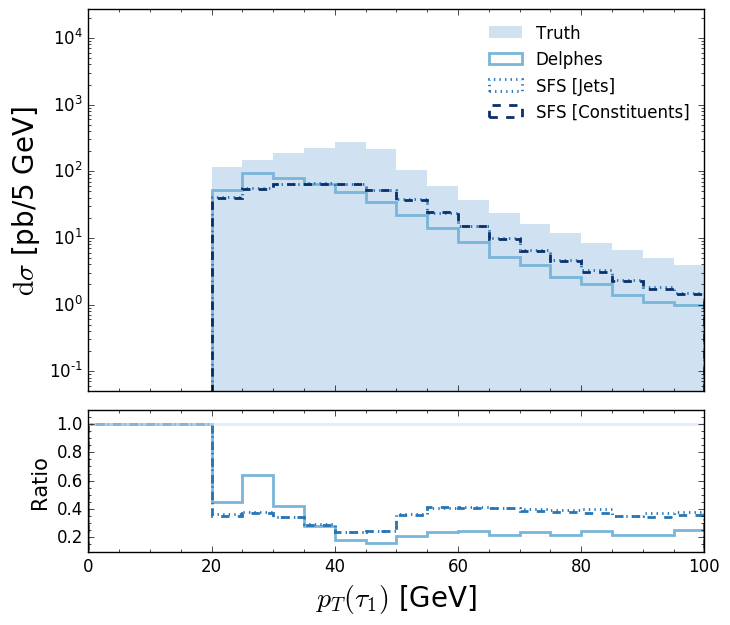}
\caption{\it Same as figure~\ref{fig:jet_props}, but for the reconstructed
  hadronic tau multiplicity distribution (upper left), and the pseudo-rapidity
  (lower left) and transverse momentum (lower right) spectrum of the leading
  tau. We additionally include the missing transverse energy spectrum (upper
  right). \label{fig:tau_props}}
\end{figure*}

This feature consequently impacts any observable, potentially more inclusive,
that would depend on the properties of the various leptons. This is illustrated
in figure~\ref{fig:dy_met} where we present the missing transverse energy
($\slashed{E}_T$) spectrum. In this case, the differences are even more
pronounced than for the lepton $p_T$ spectra, due to the particle flow method
used in \delphes\ to reconstruct the missing energy of each event. In contrast,
\madanalysis\ only sums vectorially the momentum of all visible reconstructed
objects.
Most differences occur in the low-energy region of the distribution, that is
largely impacted by the differently modelled softer objects, but the
effects are also significant for large $\slashed{E}_T$ values.

\subsection{Tau pair-production}\label{sec:tau}

We now move on with a comparison of the detector performance for the
reconstruction of hadronic taus, for which \delphes\ and our simplified detector
simulation follow quite different methods. \delphes\ uses a cone-based algorithm
to identify hadronic taus from the jet collection, whereas in the SFS approach,
hadronic taus are tagged through the matching of a reconstructed jet with a
hadron-level object. Such a matching imposes that the angular
distance, in the transverse plane, between a reconstructed tau and a
Monte Carlo hadron-level (and thus un-decayed) tau is smaller than some
threshold. Such differences
between the \delphes\ and SFS approaches can cause large variations in the
properties of the reconstructed taus, as shown in the rest of this
section.

To quantify the impact of this difference, we produce a sample of
250,000 di-tau events,
$pp\to \tau^+\tau^-$, that includes a 10~GeV selection on the tau transverse
momentum at the generator level, as well as a minimum requirement of
50~GeV on the invariant mass of the di-tau system.
In the SFS case, we reconstruct jets, electrons
and muons as described in sections~\ref{sec:QCD} and \ref{sec:DY}. We moreover
implement a tau reconstruction efficiency $\varepsilon_{\rm tracks}$ aiming at
reproducing tracker effects,
\be
  \varepsilon_{\rm tracks}(p_T) = 0.7\
    \Theta\bigg[\frac{p_T}{\rm GeV}-20\bigg] \ ,
\ee
as well as a tau tagging efficiency $\varepsilon_{\tau|\tau}$ that is identical
to the one embedded in the standard \delphes\ CMS detector parametrisation, and
that is independent of the tau transverse momentum,
\be
  \varepsilon_{\tau|\tau}(\eta) = 0.6\ \Theta\big[2.5-|\eta|\big] \ .
\label{eq:tata}\ee
The corresponding mistagging rate of a light jet as a tau $\varepsilon_{\tau|j}$
is flat and independent of the kinematics,
\be \varepsilon_{\tau|j} = 0.01 \ .\label{eq:taj}\ee
In addition, the properties of each reconstructed tau object have been smeared
as for jets (see section~\ref{sec:QCD}).

We follow the same analysis strategy as in section~\ref{sec:QCD}. 
We first select as jet and lepton candidates those jets and
leptons with a transverse momentum greater than 20~GeV and 10~GeV respectively.
Then, we remove from the jet collection any jet lying at $\Delta R_{ej} < 0.2$
of any reconstructed electron, and any electron or muon lying at
$\Delta R_{\ell j} < 0.4$ from any of the remaining jets.
In the SFS framework, the (true) hadronic tau collection is first extracted
from the event history. In contrast, in \delphes, hadronic taus are defined from
the jet collection. To account for this difference, we implement identical
selections on taus and jets at the level of the analysis.
Additionally, we restrict the collection of tau candidates and select only taus whose
transverse momentum is larger than 20~GeV, and remove any potential overlap
between the tau and the jet collection by ignoring any jet that is too
close to a tau ($\Delta R_{\tau j}<0.2$).

In figure~\ref{fig:tau_props}, we compare predictions obtained with the SFS
approach, \delphes\ and the Monte Carlo truth (\ie\ for an ideal detector).

We start by considering the tau multiplicity spectrum (upper left). As expected
from the imperfections of the tagger of eqs.~\eqref{eq:tata} and \eqref{eq:taj},
a large number of true tau objects are not tagged as such, for all three
detector simulation setups. We observe a quite good agreement between
SFS-based and \delphes-based results, despite the above-mentioned differences
between the two ap\-proa\-ches for tau tagging.

As for the electron and muon case, predictions for the pseudo-rapidity spectrum
of the leading tau (lower left) are comparable, regardless of the adopted
detector simulator. The main differences arise in the $|\eta|>2.5$
forward regime. For
SFS-based results, forward taus are stemming from the misidentification of a
light jet as a tau, whereas in the \delphes\ case, the entire jet collection
(including $c$-jets and $b$-jets) is used. This different treatment, on top of
the above-mentioned differences inherent to the whole tau-tagging method
as well as the statistical limitations of the generated event sample
in the forward regime (that actually dominate), leads
to the observed deviations between the predictions for $|\eta|>2.5$.

In the lower right panel of figure~\ref{fig:tau_props}, we present the
distribution in the transverse momentum of the leading tau. As for
the other considered observables, the detector effects are pretty important when
compared with the Monte Carlo truth. For
hard taus with $p_T\gtrsim 50$~GeV, \delphes\ and SFS-based predictions are in
very good agreement. However, the different treatment in the two simulators
significantly impacts the lower $p_T$ regime, leading in a very different
behaviour.

In the upper right panel, we show how those effects impact a more global
observable. This is illustrated with the missing transverse energy
$\slashed{E}_T$ (that is
reconstructed with a particle flow algorithm in \delphes). The discrepancies
between \delphes\ and SFS results are not so drastic as for the $p_T(\tau_1)$
distribution in the soft regime, but impact instead the entire spectrum with a
shift of ${\cal O}(10)\%$ in one way or the other. When $\slashed{E}_T >
100$~GeV (not shown on the figure), however, we enter the hard regime where the
exact details of the detector simulation matter less and a good agreement is
obtained between all predictions.

\subsection{Photon production}
\label{sec:photons}

This subsection is dedicated to the last class of objects that could be
reconstructed in a detector, namely photons. To compare the expectation from
predictions made with \delphes\ to those made with the simplified fast detector
simulator of \madanalysis, we generate a sample of 250,000
di-photon events, $pp \to 
\gamma \gamma$. In our simulations, we impose a generator level selection of
10~GeV on the photon $p_T$, as its pseudo-rapidity is constrained to be below
5 in absolute value.

\begin{figure}
  \centering
  \includegraphics[scale=0.44]{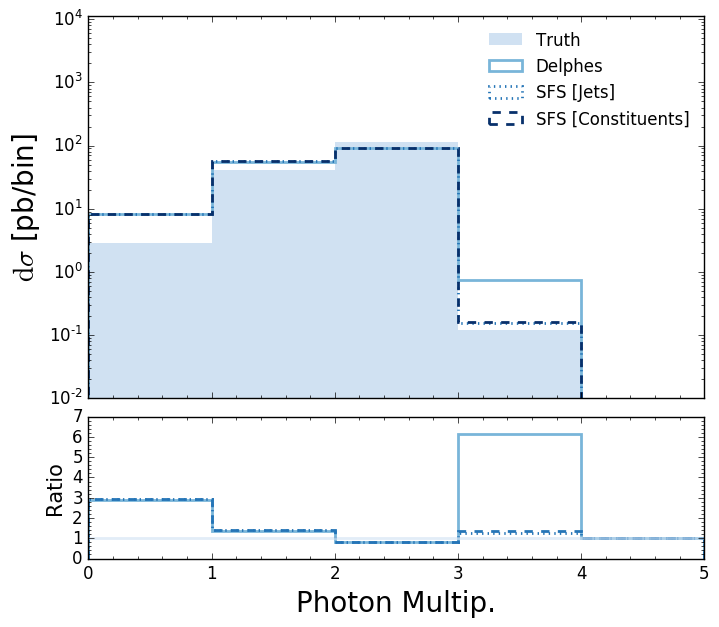}\\[.4cm]
  \includegraphics[scale=0.44]{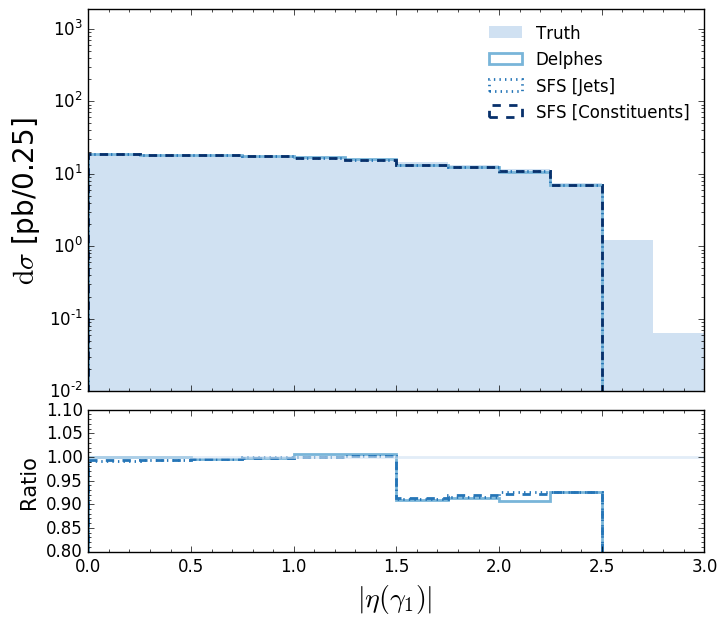}\\[.4cm]
  \includegraphics[scale=0.44]{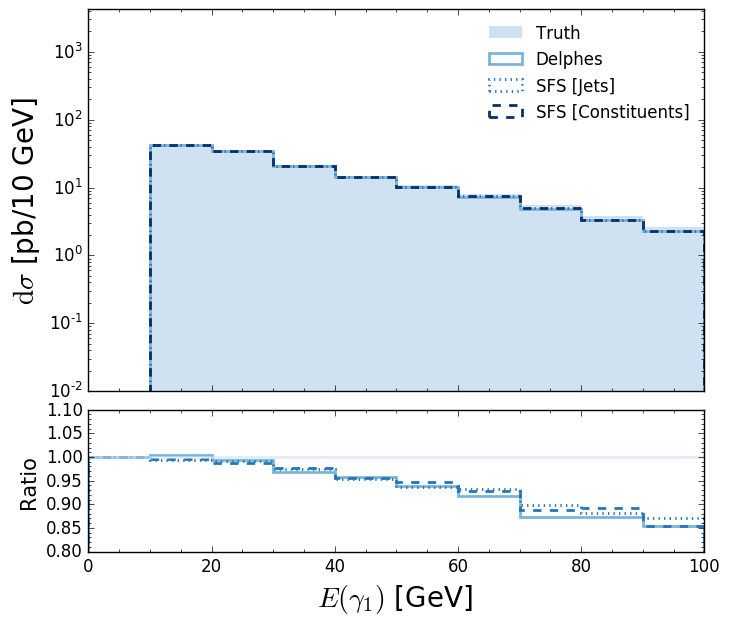}
\caption{\it Same as figure~\ref{fig:jet_props}, but for the reconstructed
  photon multiplicity distribution (upper), and for the pseudo-rapidity
  (centre) and energy (lower) spectrum of the leading
  reconstructed photon. \label{fig:a_props}}
\end{figure}

The SFS simulation includes a reconstruction efficiency $\varepsilon_\gamma$
given by
\be\renewcommand{\arraystretch}{1.4}
  \varepsilon_\gamma(p_T, \eta) =
  \left\{\begin{array}{l l}
    0.96  & \hspace{0.15cm}{\rm for~}
       p_T\!>\!10~{\rm GeV~and~} |\eta|\!\leq\! 1.5\ ,\\
    0.875  & \hspace{0.15cm}{\rm for~}
       p_T\!>\!10~{\rm GeV~and~} 1.5\!<\! |\eta| \!\leq\! 2.5\ ,\\
    0     & \hspace{0.15cm}{\rm otherwise}\ ,
  \end{array}\right.
\ee
and the photon energy is smeared as in eq.~\eqref{eq:e_smear}, this
smearing originating from the same electronic calorimeter effects as in
the electron case.

At the analysis level, we follow the same strategy as in the previous 
sections. In practice, we begin with rejecting all leptons and
photons with a transverse momentum smaller than 10~GeV, and all jets with a
transverse momentum smaller than 20~GeV. Then, we remove from the jet collection
any jet that lies at a distance in the transverse plane $\Delta R_{ej}<0.2$
from any electron candidate,
and next remove from the lepton collection any lepton that would lie at a
distance $\Delta R_{\ell j} < 0.4$ of any of the remaining jets. We finally
remove, from the photon collection, any photon lying at a distance
$\Delta R_{aj} < 0.4$ of a jet.

In order to assess the impact of the detector on the photons, we compare in
figure~\ref{fig:a_props} the Monte Carlo truth predictions (ideal detector) with
the results obtained with the \madanalysis\ SFS simulation and with \delphes.
In all cases, the detector capabilities in reconstructing the photons are quite
good, only a few photons being lost, and their properties are nicely reproduced.
Some exceptions are in order, in particular for the photon
multiplicity spectrum. Predictions using \delphes\ and those relying on the SFS
framework are quite different for bins associated with a number of photons
larger than 2.
Those extra photons (relatively to the two photons originating from the hard
process) come the hadronic decays following the hadronisation process. In some
rare cases, they are produced at wide angles and are thus not clustered back by
the jet algorithm. As photon isolation and reconstruction are treated quite
differently in the SFS framework and in \delphes,
differences are expected, as visible from the photon multiplicity figure.

\section{Reinterpreting the results of the LHC}\label{sec:pad}

\subsection{Generalities}

The LHC collaborations usually interpret their results for a specific set of
selected models. This hence leaves the task of the reinterpretation in other
theoretical frameworks to studies that have to be carried out outside the
collaborations. The most precise method that is available to theorists in this
context relies on the Monte Carlo simulation of any new physics signal of
interest. The resulting events are first reconstructed by including a detector
simulation mimicking the ATLAS or CMS detector, and next studied in order to see
to which extent the signal regions of a given analysis are populated. From those
predictions, it is then possible to conclude about the level of exclusion of the
signal, from a comparison with data and the Standard Model expectation.

Such an analysis framework is available within \madanalysis\ for half a
decade~\cite{Dumont:2014tja,Conte:2018vmg}. In practice, it makes use of the
\madanalysis\ interface to \delphes\ to handle the simulation of the detector.

In this work, we have extended this infrastructure, so that the code offers,
from version 1.8.51 onwards, the choice to employ either \delphes\ or the SFS to
deal with the simulation of the detector response. Similarly to what is done in
{\sc Rivet}~\cite{Buckley:2010ar} or {\sc ColliderBit}~\cite{Balazs:2017moi},
the simulation of the detector can now be handled together with a simple event
reconstruction to be performed with \fastjet, through transfer functions
embedding the various reconstruction and tagging efficiencies (see
section~\ref{sec:sfs}).

The estimation of the LHC potential relatively to any given new physics signal
can be achieved by typing, in the \madanalysis\ interpreter (after having started
the programme in the reconstruction-level mode),
\begin{verbatim}
 set main.recast = on
 import <events.hepmc.gz>
 submit
\end{verbatim}
The above set of commands turns on the recasting module of the platform and
allows for the reinterpretation of the results of all implemented LHC analyses
available on the user system. The signal to test is described by the
\verb+<events.hepmc.gz>+ hadron-level event sample. After submission,
\madanalysis\ generates a recasting card requiring to switch on or
off any of all implemented analyses, regardless that they
use \delphes\ or the SFS for the simulation of the detector response. On
run time, the code automatically chooses the way to handle it, so that the user
does not have to deal with it by themselves. We refer to
refs.~\cite{Conte:2018vmg,Araz:2019otb} for more information on LHC recasting in
the \madanalysis\ context, the installation of the standard Public Analysis
Database analyses (PAD) that relies on \delphes, as well as for a
detailed list with all available options.

All analyses that have been implemented and validated within the SFS context
can be downloaded from the internet and locally installed by typing, in
the command-line interface,
\begin{verbatim}
 install PADForSFS
\end{verbatim}
Up to now, this command triggers the installation of four
analyses, namely the ATLAS-SUSY-2016-07 search for glu\-i\-nos and squarks in
the multi-jet plus missing transverse energy channel~\cite{Aaboud:2017vwy}
and its ATLAS-CONF-2019-040 full run~2
update~\cite{ATLAS:2019vcq}, the ATLAS-SUSY-2018-31 search for sbottoms when
their decay gives rise to many $b$-jets (possibly originating from intermediate
Higgs boson decays) and missing transverse energy~\cite{Aad:2019pfy} and the
CMS-SUS-16-048
search for charginos and neutralinos through a signature comprised of soft
leptons and missing transverse energy~\cite{Sirunyan:2018iwl}. Details about the
validation of the SFS implementations of the ATLAS-SUSY-2016-07 and
CMS-SUS-16-048 analyses are provided in sections~\ref{sec:atlas_recast} and
\ref{sec:cms_recast} respectively. The implementation of the
ATLAS-SUY-2018-31 and its validation have been documented in
refs.~\cite{Fuks:2021wpe,Araz:2020stn}, and we
refer to the \href{http://madanalysis.irmp.ucl.ac.be/wiki/PublicAnalysisDatabase}{\madanalysis\ website}\footnote{See the webpage \url{http://madanalysis.irmp.ucl.ac.be/wiki/PublicAnalysisDatabase}.}
for information about the implementation and validation of the last
analyses. This webpage will maintain an up-to-date
list with all validated analyses available for LHC recasting with an SFS
detector simulation, in addition to those that could be used with \delphes\ as a
detector simulator.

In the future, \madanalysis\ aims to support both \delphes-based and
SFS-based LHC recasting. On the one hand, this strategy
prevents us from having to re-implement, in the SFS framework, any single
PAD analysis that is already available when relying on \delphes\ as a detector
simulator. Second, this leaves more freedom to the user who would like to
implement a new analysis in the PAD for what concerns the choice of the
treatment of
the detector effects. It should however be kept in mind that new functionalities
that are currently being developped will extend the built-in SFS capabilities of
\madanalysis, and not add any new feature to \delphes. For instance, methods
to deal with long-lived particles going
beyond what \delphes\ could do are already available from the version 1.9.10 of
\madanalysis~\cite{llp}.

In the rest of this section, we compare the SFS-based predictions with those
resulting from the usage of the \delphes\ software for the simulation of the
response of the LHC detectors for the two considered analyses.
Additionally to a direct comparison of a \delphes-based and
transfer-function-based approach for LHC recasting, this allows one to assess
the capabilities of the SFS approach as compared with \delphes\ in the case of
events featuring very hard objects that are typical of most searches for new
physics at the LHC. Very hard jets are in particular considered in
the ATLAS-SUSY-2016-07 analysis, which contrasts with the study of the jet
properties achieved in section~\ref{sec:validation} that solely covers objects
featuring a moderate transverse momentum of 10--100~GeV.

\subsection{Recasting a multi-jet plus missing energy ATLAS search for squarks
and gluinos}\label{sec:atlas_recast}

In the ATLAS-SUSY-2016-07 analysis, the ATLAS collaboration searches for squarks
and gluinos through a signature comprised of 2 to 6 jets and a large amount of
missing transverse energy. A luminosity of 36.1~fb$^{-1}$ of proton-proton
collisions at a centre-of-mass energy of 13~TeV is analysed. This search
includes two classes of signal regions. The first one relies on the effective
mass variable $M_{\rm eff}(N)$, defined as the scalar sum of the transverse
momentum of the $N$ leading jets and the missing transverse energy, and the
second one on the recursive jigsaw reconstruction technique~\cite{%
Jackson:2017gcy}. However, only the former region can be recasted due to the
lack of public information associated with the signal regions
relying on the jigsaw reconstruction technique.

Consequently, only signal regions depending on the
$M_{\rm eff}(N)$ quantity have been implemented in the \madanalysis\
framework, as detailed in ref.~\cite{Chalons:AtlasSusy1607}. In this
implementation, the simulation of the detector is handled with \delphes\ and
an appropriately tuned parameter card. This \delphes-based recast code has been
validated by reproducing public information provided by the ATLAS collaboration,
so that we use it as a reference below. In the following, we denote
predictions obtained with it as `PAD' results, the acronym PAD referring to the
traditional Public Analysis Database of \madanalysis\ that relies on \delphes\
for the simulation of the detector. In contrast, results obtained by using an
SFS detector simulation are tagged as `SFS' or `PADForSFS' results.

In order to illustrate the usage of the SFS framework for LHC recasting, we
have modified the original ATLAS-SUSY-2016-07 analysis implementation and
included it in the PADForSFS database, together with an appropriate ATLAS
SFS detector parametrisation. The user has thus the choice to use either
\delphes\ or the SFS framework for the reinterpretation of the results of this
ATLAS analysis.

The validation of our implementation has been a\-chie\-ved by comparing
predictions for a well-defined benchmark scenario with those obtained with the
reference version of the implementation based on \delphes. We have adopted a
simplified model setup inspired by the Minimal Supersymmetric Standard Model in
which all superpartners are decoupled, with the exception of the gluino and the
lightest neutralino. Their masses have been fixed to 1~TeV and 825~GeV
respectively, and the gluino is enforced to decay
into a di-jet plus neutralino system with a branching ratio of 1.

We have generated 200,000 new physics Monte Carlo events by matching
leading-order
matrix elements convoluted with the leading order set of NNPDF~2.3 parton
densities~\cite{Ball:2012cx} as generated by \madgraph~\cite{Alwall:2014hca},
with the parton shower machinery of \pythia~\cite{Sjostrand:2014zea}. Gluino
decays are handled with the {\sc MadSpin}~\cite{Artoisenet:2012st} and
{\sc MadWidth}~\cite{Alwall:2014bza} packages, and we have relied on \pythia\
for the simulation of the hadronisation
processes. Those events have then been analysed automatically in \madanalysis,
both in the `PAD' context with a \delphes-based detector simulation and in the
SFS context with an SFS-based detector simulation.

We have compared the two sets of results and found that they deviate by at most
10\% for all signal regions populated by at least 20 events (out of the 200,000
simulated events). In addition, we have verified that enforcing a
constituent-based or jet-based jet smearing had little impact on the results. We
have observed that this choice indeed leads to a modification of the SFS
predictions of about 1\%, so that the jet-smearing choice
is irrelevant. Jet-based jet smearing is therefore used below. Furthermore,
in terms of performance, the run of \delphes\ has been found 50\% slower than
the SFS one.

\begin{table}
  \centering
  \renewcommand{\arraystretch}{1.4}
  \setlength\tabcolsep{3pt}
  \begin{tabular}{c|cc|cc|c}
    Cuts & $n_i^{\rm PAD}$& $\varepsilon_i^{\rm PAD}$ & $n_i^{\rm SFS}$ &
     $\varepsilon_i^{\rm SFS}$ & $\delta_i$\\ \hline
      Initial         & 7262.2 & -      & 7262.2 & -      & -      \\
      Preselection    & 1129.8 & 0.156 & 1092.0 & 0.150 & 3.3\% \\
      At least 2 jets & 1125.0 & 0.996 & 1087.5 & 0.996 & 0 \\
      At least 3 jets & 1073.0 & 0.954 & 1032.9 & 0.950 & 0.4\% \\
      $\Delta\varphi(j_k, \vec{\slashed{p}_T}) > 0.4$
                      &  848.5 & 0.791 &  821.6 & 0.796 & 0.6\% \\
      $\Delta\varphi(j_3, \vec{\slashed{p}_T}) > 0.2$
                      &  767.6 & 0.905 &  743.9 & 0.905 & 0.1\% \\
      $p_T(j_1)>700$~GeV
                      &   53.9 & 0.070 &   54.8 & 0.074 & 4.9\% \\
      $p_T(j_2)>50$~GeV
                      &   53.9 & 1     &   54.8 &1 &  0 \\
      $p_T(j_3)>50$~GeV
                      &   53.9 & 1     &  54.8 &1 &  0\\
      $\slashed{E}_T/\sqrt{H_T} > 16~\sqrt{\rm GeV}$
                      &   43.0 &  0.798 &  44.4 & 0.812 & 1.7\% \\
      $M_{\rm eff}(3) > 1300$~GeV & 43.0 & 1 &44.4 & 1 & 0\\[.05cm]
      \multicolumn{6}{c}{}\\[.05cm]
    Cuts & $n_i^{\rm PAD}$& $\varepsilon_i^{\rm PAD}$ & $n_i^{\rm SFS}$ &
     $\varepsilon_i^{\rm SFS}$ & $\delta_i$\\ \hline
      Initial & 7262.2 & - &7262.2 & - & -\\
      Preselection    & 1129.8 & 0.156 &1092.0 & 0.150 & 3.3\%\\
      At least 2 jets & 1125.0 & 0.996 &1087.5 & 0.996 & 0\\
      At least 4 jets &  857.4 & 0.762 & 821.2 & 0.755 & 0.9\%\\
      $\Delta\varphi(j_k, \vec{\slashed{p}_T}) > 0.4$
                      &  677.1 & 0.790 & 652.3 & 0.794 & 0.6\%\\
       $\Delta\varphi(j_l, \vec{\slashed{p}_T}) > 0.4$
                      &  522.0 & 0.771 & 500.8 & 0.768 & 0.4\%\\
       $p_T(j_4)>100$~GeV & 113.5 & 0.218 &112.5 & 0.225 & 3.3\%\\
       $|\eta(j)| < 2$ & 91.1 & 0.803\% &89.0 & 0.791 & 1.5\%\\
       Aplanarity $> 0.04$ & 54.5 & 0.598 &52.9 & 0.594 & 0.7\%\\
      $\slashed{E}_T/M_{\rm eff}(4) > 0.25$
                      & 49.2 & 0.902 &48.2 & 0.911 & 1.0\%\\
        $M_{\rm eff}(4)> 1400$~GeV & 27.0 & 0.549 &27.5 & 0.570 & 3.8\%\\[.05cm]
      \multicolumn{6}{c}{}\\[.05cm]
    Cuts & $n_i^{\rm PAD}$& $\varepsilon_i^{\rm PAD}$ & $n_i^{\rm SFS}$ &
     $\varepsilon_i^{\rm SFS}$ & $\delta_i$ \\ \hline
      Initial & 7262.2 & - &7262.2 & - & -\\
      Preselection & 1129.8 & 0.156 &1092.0 & 0.15 & 3.3\%\\
       At least 2 jets & 1125.0 & 0.996 &1087.5 & 0.996 & 0\\
       At least 6 jets & 244.2 &  0.217 &225.2 & 0.207 & 4.6\%\\
      $\Delta\varphi(j_k, \vec{\slashed{p}_T}) > 0.4$
           & 194.0 & 0.795 &179.5 & 0.797 & 0.3\%\\
      $\Delta\varphi(j_l, \vec{\slashed{p}_T}) > 0.2$
           & 154.9 & 0.798 &143.3 & 0.799 & 0\\
      $|\eta(j)| < 2$ & 101.4 & 0.655 &93.0 & 0.649 & 0.8\%\\
      $\slashed{E}_T/M_{\rm eff}(6) > 0.25$ 
           & 85.9 & 0.847 &78.2 & 0.841 & 0.8\%\\
      $M_{\rm eff}(6) > 1200$~GeV & 67.6 & 0.787 &61.6 & 0.788 & 0.1\%
    \end{tabular}
    \caption{\it Cut-flow charts associated with three signal regions of the
      ATLAS-SUSY-2016-07 analysis, among the most populated ones by a 1~TeV
      gluino signal. We focus on regions relevant for events featuring at
      least 3 jets (top), 4 jets (middle) or 6 jets (lower). In our notation,
      $j_k$ denotes any of the three leading jets, and $j_l$ any jet from the
      third one. \label{tab:atlas_recast}}
\end{table}

We present a subset of the results in table~\ref{tab:atlas_recast}, focusing on
three of the ATLAS-SUSY-2016-07 signal regions that are among the most populated
ones by the considered 1~TeV gluino signal. We show predictions obtained by
using \delphes\ for the simulation of the detector (PAD), and depict them both
in terms of the number of events $n_i$ surviving a cut $i$ and of the related
cut efficiency
\be \varepsilon_i = n_i/n_{i-1}\ .\ee
We additionally display SFS-based predictions (SFS),
showing again both the number of events and the various cut efficiencies.

The deviations $\delta_i$ between the \delphes\ and SFS predictions are
evaluated at the level of the efficiencies,
\be
  \delta_i = \bigg|1-
    \frac{\varepsilon_i^{\rm SFS}}{\varepsilon_i^{\rm PAD}}\bigg|\ .
\ee
As above-mentioned, the deviations at any cut-level are smaller than 10\%, and
often lie at the level of 1\%. This is in particular the case for the cuts
relevant to the other 19 (not shown) signal regions. A more complete set of
results, including predictions for all
signal regions, is available online\footnote{See the PADForSFS webpage
\url{http://madanalysis.irmp.ucl.ac.be/wiki/SFS}.}.

\subsection{Recasting a CMS search for compressed electroweakinos with soft
leptons and missing energy} \label{sec:cms_recast}

The CMS-SUS-16-048 analysis is an unusual search for the supersymmetric partners
of the Standard Model gauge and Higgs bosons. It relies on the reconstruction of
soft leptons with a transverse momentum smaller than 30~GeV, and mainly targets
the associated production of a neutralino and chargino pair $\tilde\chi^\pm_1
\tilde\chi_2^0$ in a setup in which the supersymmetric spectrum is compressed
and features small mass splittings between these two states and the lightest
neutralino $\tilde\chi_1^0$. This search investigates 35.9~fb$^{-1}$ of
proton-proton collisions at a centre-of-mass energy of 13~TeV.

The production process under consideration ($p p \to \tilde\chi^\pm_1
\tilde\chi_2^0$) is thus followed by a decay of both superpartners into the
lightest neutralino $\tilde\chi_1^0$ and an off-shell gauge boson,
\be
 p p \to \tilde\chi^\pm_1\tilde\chi_2^0
    \to (W^\ast \tilde\chi_1^0)\ (Z^\ast \tilde\chi_1^0) \ .
\label{eq:cms_signal}\ee
The considered signal is thus potentially comprised of a pair of opposite-sign
(OS) soft leptons (arising from the off-shell gauge bosons), possibly carrying
the same flavour (SF), and some missing
transverse momentum carried away by the two produced lightest neutralinos.

The CMS collaboration has also designed a series of signal regions
dedicated to the search for electroweakino production from stop decays,
\be
 p p \to \tilde t \tilde t^* \to (\chi^\pm_1 b)\ (\chi^\mp_1 \bar b)
    \to (W^\ast \tilde\chi_1^0 b)\ (W^\ast \tilde\chi_1^0 \bar b) \ .
\ee
As for the process of eq.~\eqref{eq:cms_signal}, such a signal also gives rise
to soft leptons, produced this time in association with $b$-jets. The top squark
is however compressed with the other states, so that those $b$-jets are in most
cases not identified. The resulting signature is therefore very similar to the
one originating from eq.~\eqref{eq:cms_signal}.

\begin{table}
  \centering
  \renewcommand{\arraystretch}{1.38}
  \setlength\tabcolsep{4pt}
    \begin{tabular}{c|cc|cc|c}
    Cuts & $n_i^{\rm PAD}$& $\varepsilon_i^{\rm PAD}$ & $n_i^{\rm SFS}$ &
     $\varepsilon_i^{\rm SFS}$ & $\delta_i$\\ \hline
    Initial                        & 48572.7 & - & 48572.7 & - & - \\
    At least 1 lepton              & 17028.2 & 0.351 &17369.5 & 0.358 & 2.0\%\\
    $p_T(\ell_1)\in[5, 30]$~GeV    & 13142.0 & 0.772 &13266.2 & 0.764 & 1.0\%\\
    $p_T(\ell_2)<30$~GeV           &  1677.8 & 0.128 & 1688.4 & 0.127 & 0.3\%\\
    OS dilepton                    &  1042.6 & 0.621 & 1041.4 & 0.617 & 0.7\%\\
    $p_T(\ell_1\ell_2)>3$~GeV      &   961.6 & 0.922 &  963.9 & 0.926 & 0.4\%\\
    $\slashed{E}_T > 125$~GeV      &   148.9 & 0.155 &  156.2 & 0.162 & 4.6\%\\
    $\slashed{E}^{\rm cor.}_T>125$~GeV
                                   &    90.5 & 0.608 &   93.2 & 0.597 & 1.9\%\\
    $\slashed{E}_T > 300$~GeV      &    27.9 & 0.308 &   25.8 & 0.277 & 10.0\%\\
    At least 1 jet                 &    27.8 & 0.997 &   25.7 & 0.994 & 0.3\%\\
    $H_T>100$~GeV                  &    27.2 & 0.979 &   25.0 & 0.974 & 0.5\%\\
    $\slashed{E}_T/H_T\in[0.6, 1.4]$
                                   &    25.2 & 0.924 &   23.7 & 0.948 & 2.5\%\\
    $b$-jet veto                   &    15.3 & 0.609 &   14.6 & 0.616 & 1.0\%\\
    $M_{\tau\tau}\not\in[0,160]$~GeV
                                   &    12.1 & 0.788 &   11.7 & 0.799 & 1.5\%\\
    $p_T(\ell_1)<12$~GeV           &     5.1 & 0.426 &    5.7 & 0.484 & 13.7\%
    \end{tabular}\\[.2cm]
    \begin{tabular}{c|cc|cc|c}
    Cuts & $n_i^{\rm PAD}$& $\varepsilon_i^{\rm PAD}$ & $n_i^{\rm SFS}$ &
     $\varepsilon_i^{\rm SFS}$ & $\delta_i$\\ \hline
    Initial                        & 48572.7 & -     &48572.7 & -     & - \\
    At least 1 lepton              & 17028.2 & 0.351 &17369.5 & 0.358 & 2.0\%\\
    $p_T(\ell_1)\in[5, 30]$~GeV    & 13142.0 & 0.772 &13266.2 & 0.764 & 1.0\%\\
    $p_T(\ell_2)<30$~GeV           &  1677.8 & 0.128 & 1688.4 & 0.127 & 0.3\%\\
    $p_T(\ell_2)>5$~GeV            &   922.7 & 0.550 &  923.1 & 0.547 & 0.6\%\\
    SF dilepton                    &   597.8 & 0.648 &  600.0 & 0.650 & 0.3\%\\
    OS dilepton                    &   360.2 & 0.603 &  359.0 & 0.598 & 0.7\%\\
    $p_T(\ell_1\ell_2)>3$~GeV      &   331.4 & 0.920 &  328.3 & 0.914 & 0.6\%\\
    $M_{\ell_1\ell_2}\in[4, 50]$~GeV
                                   &   245.0 & 0.739 &  245.6 & 0.748 & 1.2\%\\
    Hadronic res.~veto             &   237.8 & 0.971 &  238.1 & 0.969 & 0.1\%\\
    $\slashed{E}_T > 125$~GeV      &    40.0 & 0.168 &   44.6 & 0.187 &11.5\%\\
    $\slashed{E}^{\rm cor.}_T>125$~GeV
                                   &    26.1 & 0.654 &   29.1 & 0.652 & 0.3\%\\
     $\slashed{E}_T > 250$~GeV     &    13.1 & 0.500 &   14.5 & 0.499 & 0.1\%\\
    At least 1 jet                 &    12.8 & 0.981 &   14.3 & 0.983     & 0.2\%\\
    $H_T>100$~GeV                  &    12.7 & 0.987 &   13.9 & 0.977 & 1.1\%\\
    $\slashed{E}_T/H_T\in[0.6, 1.4]$
                                   &    11.3 & 0.889 &   13.0 & 0.929 & 4.4\%\\
    $b$-jet veto                   &     7.5 & 0.666 &    8.9 & 0.684 & 2.7\%\\
    $M_{\tau\tau}\not\in[0,160]$~GeV
                                   &     5.5 & 0.738 &    7.0 & 0.788 & 6.8\%\\
    $M_T(\ell, \vec{\slashed{p}}_T)\!<\!70$~GeV
                                   &     4.5 & 0.821 &    6.1 & 0.871 & 6.1\%\\
    $M_{\ell_1\ell_2}<9$~GeV       &     3.8 & 0.837 &    4.8 & 0.785 & 6.2\%
    \end{tabular}
    \caption{\it Cut-flow charts associated with two signal regions of the
      CMS-SUS-16-048 analysis, among the most populated ones by the considered
      compressed electroweakino scenario (see the text). We focus on a region
      dedicated to the search for electroweakino production from stop decays
      (upper) and a
      region dedicated to direct electroweakino production (lower). In our
      notation, $\slashed{E}^{\rm cor.}_T$ denotes the transverse momentum
      resulting from the vector sum of the missing momentum and the momenta of
      the two leading leptons, and $M_{\tau\tau}$ represents the invariant mass
      of the di-tau sytem that would stem from considering the two leptons as
      originating from tau decays.\label{tab:cms_recast}}
\end{table}

This CMS-SUS-16-048 analysis has been implemented in the \madanalysis\ framework
and validated in the context of the last Les Houches workshop on TeV
colliders~\cite{Brooijmans:2020yij}, both for a simulation of the detector
effects relying on \delphes\ and on the SFS infrastructure. This analysis is
therefore both included in the standard PAD and new PADForSFS database.

As in section~\ref{sec:atlas_recast}, the recast code based on \delphes\ has
been first validated against publicly available material provided by the CMS
collaboration. The SFS-based version of the analysis has been developed next,
using the \delphes-based one as a reference for the validation procedure.
Whereas both implementations were validated in ref.~\cite{Brooijmans:2020yij},
we report in the following on the comparison of \delphes-based and
SFS-based predictions for a given signal, as this fits in the spirit of this
work aiming at documenting the performance of the new SFS module of
\madanalysis\ and comparing with a more standard \delphes-based detector
simulation.

For our comparison, we again consider a simplified model inspired by the Minimal
Supersymmetric Standard Model. The Standard Model is extended by the three
electroweakino states relevant for
the signal of eq.~\eqref{eq:cms_signal}. We make use of one of the next-to-minimal
simplified scenarios proposed in ref.~\cite{Fuks:2017rio} and that is studied in
the considered CMS analysis, and set the mass of the lightest
neutralino to 142.5~GeV and the ones of the other states to 150~GeV.

We generate
300,000 signal events in the same simulation chain as the one described in the
previous section. When comparing the recasting results, we observe a good
agreement between \delphes-based and SFS-based predictions. Ignoring signal
regions featuring a poor statistics (less than 10 events out of the 300,000
generated ones), the cut efficiencies are found to differ by at most 50\%,
although the deviations are found to lie at the percent level for a large
majority of cuts from all signal regions. In addition,
using jet-based or constituent-based jet smearing does not change the
conclusions, the deviations $\delta_i$ from \delphes\ being impacted by about
1\%.

We illustrate our results in table~\ref{tab:cms_recast} for two of the
CMS-SUS-16-048 signal regions that are among the most populated ones by the
considered compressed electroweakino signal. The complete set of results
can be obtained online, on the SFS wiki page.

\section{Conclusions}\label{sec:conclusion}
We have discussed the implementation of a new detector emulator in the
\madanalysis\ framework. This extends the capacities of the platform when
\fastjet\ is used for event reconstruction.

Our work includes two
components. First, the {\sc Python}-like interpreter of \madanalysis\ has been
extended by new commands. Those allow
for an intuitive parametrisation of a high-energy physics detector through
user-defined
smearing, reconstruction and identification/mistagging efficiency functions.
Next, the C++ core of the programme has been modified so that it could handle
those user-defined functions, that are converted from
the {\sc Python}-like \madanalysis\ metalanguage to C++ by the interpreter.

On run time, \madanalysis\ automatically generates a full C++ code that
allows for event reconstruction with \fastjet\ and that incorporates the
input detector effects. The code is then automatically compiled and run.

Correspondingly, we have extended the recasting infrastructure of \madanalysis.
The user has now the option to rely on our fast simplified detector simulator
instead of \delphes\ to simulate the ATLAS and CMS detectors. Four
LHC analyses
are currently available and have been validated. Moreover, the
programme is shipped with two configuration scripts reproducing the response of
the ATLAS and CMS detectors.

In order to validate our implementation, we have compared SFS-based
predictions with
those obtained when the \delphes\ package is used instead for the simulation of
the detector effects. We have considered Standard Model processes as well as new
physics signals, the latter being investigated in the context of LHC recasting.
In all our comparisons, we have found that the variations between the two
approaches are most of the time of about 10\%. However, there are
cases where larger discrepancies are found, as could be expected by virtue of
the different strategies followed by the \delphes\ detector simulator and the
SFS approach. The main advantages of our emulator is that it allows the user to
deal with any custom detector in a very simple and user-friendly manner, either
through intuitive {\sc Python}-like commands to be cast in the \madanalysis\
interpreter, or directly by programming within the expert mode of the platform.
In addition, our simplified detector simulation is generally faster than
\delphes, as it is more light-weight. With such a machinery at hand, it becomes
thus possible to use the
code to single out a particular detector effect and hence determine the
performance of any given (future) detector.

With this work, we have hence augmented the capabilities of the \madanalysis\
package, offering a new fast and efficient way to emulate the response of a
typical high-energy physics detector. The user is now allowed to
choose between a \delphes-based or a transfer-function-based simulation of the
detector response, both methods being implemented through an intuitive set of
user-friendly methods included in the \madanalysis\ command-line interface.
Predictions compatible with those obtained when \delphes\ is used have been
found, the difference lying at the level of the uncertainties inherent to a fast
detector simulation carried out outside the experimental collaborations.

In a future release, we plan to further extend both our package and the normal
mode of \madanalysis\ to include the handling of vertex position and impact
parameter variables. Those heavier changes will consequently allow for an
appropriate modelling of long-lived particles in the context of our simplified
fast detector simulation~\cite{llp}.

\section{Acknowledgements}
The authors are grateful to Eric Conte, Mariana Frank and Sabine Kraml for
useful comments on the manuscript. JYA acknowledges the hospitality of the LPTHE
(Sorbonne University), IPPP
(Durham University) and LNF (INFN) where parts of this work have been completed
during his visits. JYA has received funding from the European Union's Horizon
2020 research and innovation programme as part of the Marie Sklodowska-Curie
Innovative Training Network MCnetITN3 (grant agreement no.~722104).

\bibliography{MA5}

\end{document}